\documentclass[10pt,twocolumn,english,aps,manuscript,final,5p,times]{revtex4}
\usepackage[T1]{fontenc}
\usepackage[latin9]{inputenc}
\usepackage{amsmath}
\usepackage{amssymb}
\usepackage{graphicx}
\usepackage{color,soul}
\makeatletter
\@ifundefined{textcolor}{}
{%
 \definecolor{BLACK}{gray}{0}
 \definecolor{WHITE}{gray}{1}
 \definecolor{RED}{rgb}{1,0,0}
 \definecolor{GREEN}{rgb}{0,1,0}
 \definecolor{BLUE}{rgb}{0,0,1}
 \definecolor{CYAN}{cmyk}{1,0,0,0}
 \definecolor{MAGENTA}{cmyk}{0,1,0,0}
 \definecolor{YELLOW}{cmyk}{0,0,1,0}
}

\include{shadowtext}

\makeatother
\pdfoutput=1
\usepackage{babel}
\begin{document}

\title{Pairwise thermal entanglement in Ising-XYZ diamond chain structure
in an external magnetic field}

\author{J. Torrico$^{1}$, M. Rojas$^{1}$, S. M. de Souza$^{1}$, Onofre Rojas$^{1}$ and N. S. Ananikyan$^{2}$}

\affiliation{$^{1}$Departamento de Ciencias Exatas, Universidade Federal de Lavras,
37200-000, Lavras-MG, Brazil}

\affiliation{$^{2}$A.I. Alikhanyan National Science Laboratory, 0036 Yerevan,
Armenia.}
\begin{abstract}
Quantum entanglement is one of the most fascinating types of correlation that can be shared only among quantum systems. The Heisenberg chain is one of the simplest quantum chains which exhibits a reach entanglement feature, due to the Heisenberg interaction is quantum coupling in the spin system. The two particles were coupled trough XYZ coupling or simply called as two-qubit XYZ spin, which are the responsible for the emergence of thermal entanglement. These two-qubit operators are bonded to two nodal Ising spins, and this process is repeated infinitely resulting in a diamond chain structure. We will discuss two-qubit thermal entanglement effect on Ising-XYZ diamond chain structure. The concurrence could be obtained straightforwardly in terms of two-qubit density operator elements, using this result, we study the thermal entanglement, as well as the threshold temperature where entangled state vanishes. The present model displays a quite unusual concurrence behavior, such as, the boundary of two entangled regions becomes a disentangled region, this is intrinsically related to the XY-anisotropy in the Heisenberg coupling. Despite a similar property had been found for only two-qubit, here we show in the case of a diamond chain structure, which reasonably represents real materials.
\end{abstract}
\maketitle

\section{Introduction}

In the last decade, many efforts were dedicated to characterizing
qualitatively and quantitatively the entanglement properties of condensed
matter systems which are the natural candidate to apply for quantum
communication, as well as quantum information. In this sense, it is
quite interesting to study the entanglement of solid state systems
such as spin chains\cite{qubit-Heisnb}. The Heisenberg chain is one of the simplest quantum chains which exhibit a reach entanglement feature, due to the Heisenberg interaction is a nonlocal correlation between quantum systems\cite{kamta}.

Quantum entanglement, with its applications to quantum phase transitions
of strongly correlated spin systems and its experimental implementation
in optical lattices, was considered, in particular, for one-dimensional
systems. Diverging entanglement length without quantum phase transition
was found in a localizable entanglement (LE) for valence bond solids
(VBSs), since the correlation length remains finite\cite{vestraete}.
This aforementioned is a rather new and remarkable result regarding the entanglement
properties of VBS quantum spin ground states. A theory for localizable
entanglement was developed based on matrix product states from the
density matrix renormalization group (DMRG) method and applied to
VBS states\cite{popp}. In reference \cite{garcia}, an experimental
implementation was proposed for VBSs of the spin-1 Heisenberg Hamiltonians
and ladders, and a method was proposed directly to measure quantum
observables that are not accessible in standard materials in condensed
matter.

Motivated by real materials such as $\mathrm{Cu_{3}(CO_{3})_{2}(OH)_{2}}$
known as azurite, which is an interesting quantum antiferromagnetic
model described by Heisenberg model on generalized diamond chain.
Honecker et al. \cite{honecker} studied the dynamic and thermodynamic
properties for this model. Moreover, the thermodynamics of the Ising-Heisenberg
model on diamond-like chain was also widely discussed in references\cite{canova06,vadim,valverde,lisnii-11}.
The motivation to research the Ising-XYZ diamond chain model is based
in some recent works. According to the experiments of the natural
mineral azurite, theoretical calculations of Ising-XXZ model, as well
as the experimental result of the exchange dimer (interstitial sites)
parameter and their descriptions of the various theoretical models.
The 1/3 magnetization plateau, the double peaks both in the magnetic
susceptibility and specific heat, was observed in the experimental
measurements\cite{rule,kikuchi}. It should be noted that the dimer
(interstitial sites) are exchange much more strong than those nodal
sites. Since dimmer interaction is much higher than the rest, it can
be represented as an exactly solvable Ising-Heisenberg model. In addition,
experimental data on the magnetization plateau coincides approximately
with Ising-Heisenberg model\cite{canova06,ananikian,chakh}. 

Recently several investigation are focusing on thermal entanglement
with Heisenberg coupling qubits as well as assuming some finite chain
structure. The thermal entanglement of isotropic Heisenberg spin chain
has been studied in the absence \cite{wang} and the presence of
an external magnetic field \cite{X-Wang,arnesen}. The entanglement
of the two-qubit isotropic Heisenberg system decreases when the temperature
increases which vanishes beyond a threshold temperature $T_{th}$.
However, two qubits with XYZ coupling displays quite interesting thermal
entanglement behavior\cite{G. Rigolin}, such as more than one threshold
temperature. On the other hand, an unusual property of entanglement
also was considered in an alternating Ising and Heisenberg spins in
simple one-dimensional chain\cite{m-rojas14}, where despite at zero
temperature there is no any evidence of entanglement, then rise a
small amount of concurrence indicating the system has a thermal entanglement
between two Heisenberg spins.  More over this result was confirmed by theoretical model \cite{lazaryan} by Gibbs-Bogoliubov approach (Heisenberg-Ising model) with experimental results  of  natural material azurite \cite{kikuchi}. 

This paper is organized as follow: in Sec. II we present the Ising-XYZ
model on diamond chain and its phase diagram at zero temperature.
Further, in Sec. III, we present the exact thermodynamic solution
of the model. In Sec. IV, we have discussed the thermal entanglement
of Heisenberg reduced density operator of the model, such as concurrence
and threshold temperature. Finally, Sec. V contains the concluding
remarks.

\section{The Ising-XYZ chain on diamond chain structure}
\begin{figure}
\includegraphics[scale=0.4]{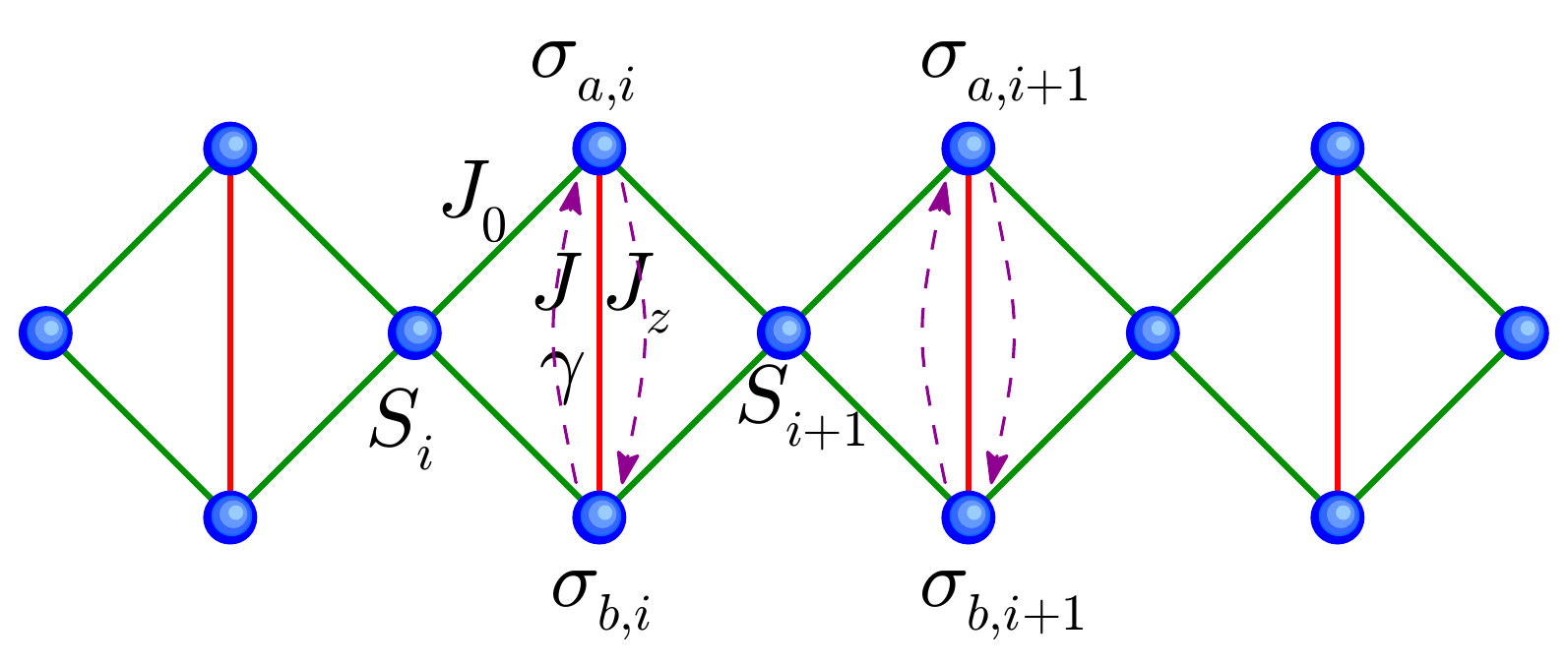}\protect\caption{\label{fig:Schem}(Color Online) Schematic representation of Ising-XYZ
chain on diamond structure, $\sigma_{a,i}$ and $\sigma_{i,b}$ are
Heisenberg spins, while $S_{i}$ corresponds to Ising spins.}
\end{figure}

 The thermal entanglement of Ising-Heisenberg diamond chain already
was discussed. Here we extend this model in agreement with the motivation
discussed above. Therefore, let us consider an Ising-XYZ diamond chain
structure as illustrated schematically in figure \ref{fig:Schem}.
Thus, the Ising-XYZ Hamiltonian becomes
\begin{alignat}{1}
H= & -\sum_{i=1}^{N}\left[J(1+\gamma)\sigma_{a,i}^{x}\sigma_{b,i}^{x}+J(1-\gamma)\sigma_{a,i}^{y}\sigma_{b,i}^{y}+\right.\nonumber \\
 & +J_{z}\sigma_{a,i}^{z}\sigma_{b,i}^{z}+J_{0}(\sigma_{a,i}^{z}+\sigma_{b,i}^{z})(S_{i}+S_{i+1})+\nonumber \\
 & \left.+h(\sigma_{a,i}^{z}+\sigma_{b,i}^{z})+\frac{h}{2}(S_{i}+S_{i+1})\right],\label{eq:Hamt}
\end{alignat}
 where $\sigma_{a(b)}^{\alpha}$ are the Pauli matrix with $\alpha=\{x,y,z\}$,
and $S$ corresponds to the Ising spins, whereas $\gamma$ is the
XY-anisotropy parameter.

After diagonalizing the XYZ term, we have the following eigenvalues for XYZ dimer 
in terms of nodal spin chain $\mu=S_{i}+S_{i+1}$,
\begin{alignat}{1}
\varepsilon_{1,4}= & -h\frac{\mu}{2}-\frac{J_{z}}{4}\pm\Delta(\mu),\\
\varepsilon_{2,3}= & -h\frac{\mu}{2}\mp\frac{J}{2}+\frac{J_{z}}{4},
\end{alignat}
wherein 
\begin{equation}
\Delta(\mu)=\sqrt{\left(h+J_{0}\mu\right)^{2}+\tfrac{1}{4}J^{2}\gamma^{2}},
\end{equation} 
with the corresponding eigenvectors in terms of standard basis $\{|\begin{smallmatrix}-\\
-
\end{smallmatrix}\rangle,|\begin{smallmatrix}-\\
+
\end{smallmatrix}\rangle,|\begin{smallmatrix}+\\
-
\end{smallmatrix}\rangle,|\begin{smallmatrix}+\\
+
\end{smallmatrix}\rangle\}$ are given respectively by 
\begin{alignat}{1}
|\varphi_{1,4}\rangle & =N_{\pm}\left(\alpha_{\pm}|\begin{smallmatrix}+\\
+
\end{smallmatrix}\rangle+|\begin{smallmatrix}-\\
-
\end{smallmatrix}\rangle\right),\\
|\varphi_{2,3}\rangle & =\tfrac{1}{\sqrt{2}}\left(|\begin{smallmatrix}-\\
+
\end{smallmatrix}\rangle\pm|\begin{smallmatrix}+\\
-
\end{smallmatrix}\rangle\right),
\end{alignat}
where $\alpha_{\pm}=\frac{-J\gamma}{2h+2J_{0}\mu\pm2\Delta(\mu)}$,
and $N_{\pm}=\frac{1}{\sqrt{1+\alpha_{\pm}^{2}}}$.

It is important to recall that the pairwise entanglement between Heisenberg
spins at zero temperature is entangled for any Hamiltonian parameters,
unless for the limiting case becomes disentangled region. Which will
be illustrated in the zero temperature limit of the phase transition. 

This model is somewhat opposite to that proposed in reference \cite{m-rojas14},
where at zero temperature there was no thermal entanglement.

\subsection{Phase diagram of Ising-XYZ on diamond chain}

\begin{figure}
\includegraphics[scale=0.2]{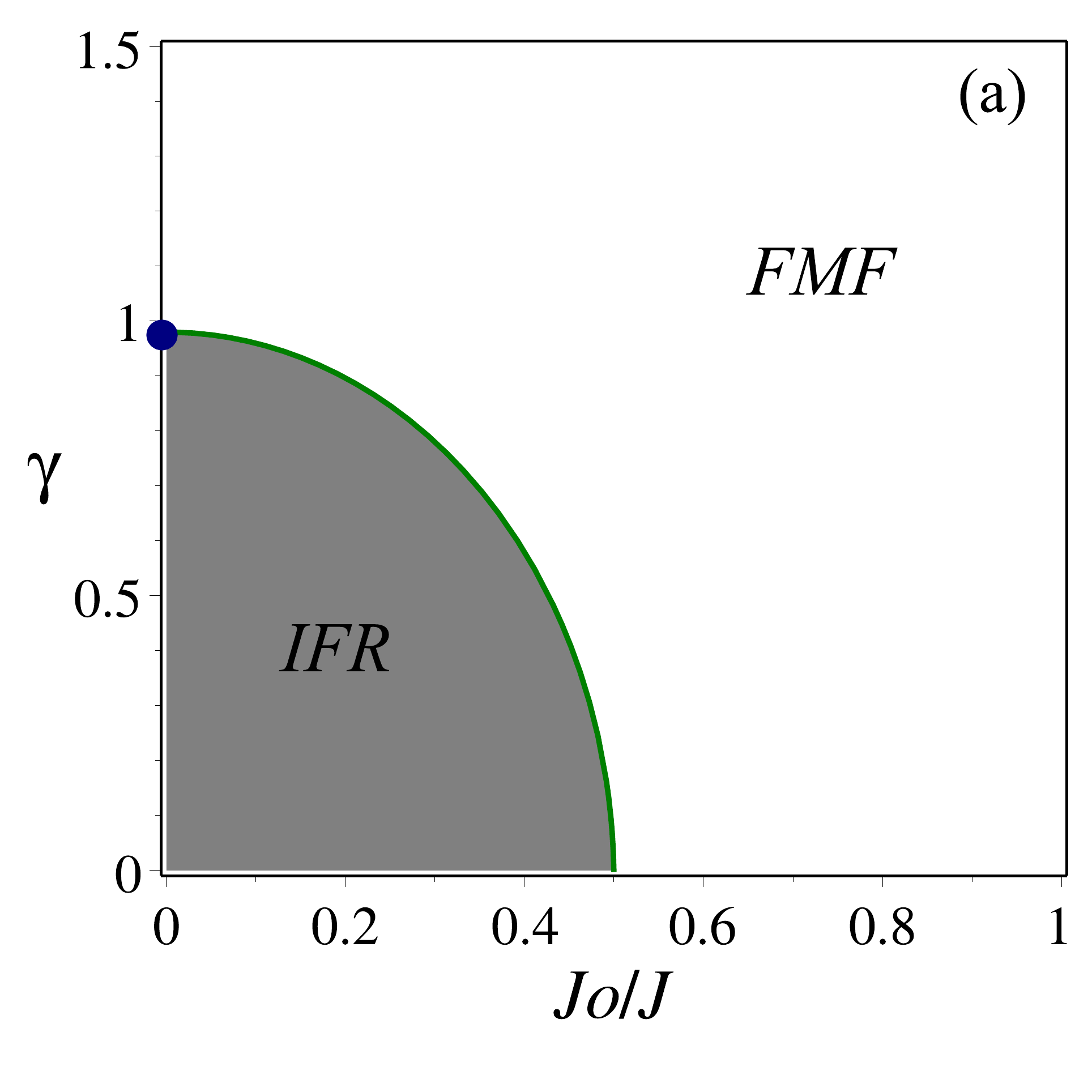}\includegraphics[scale=0.2]{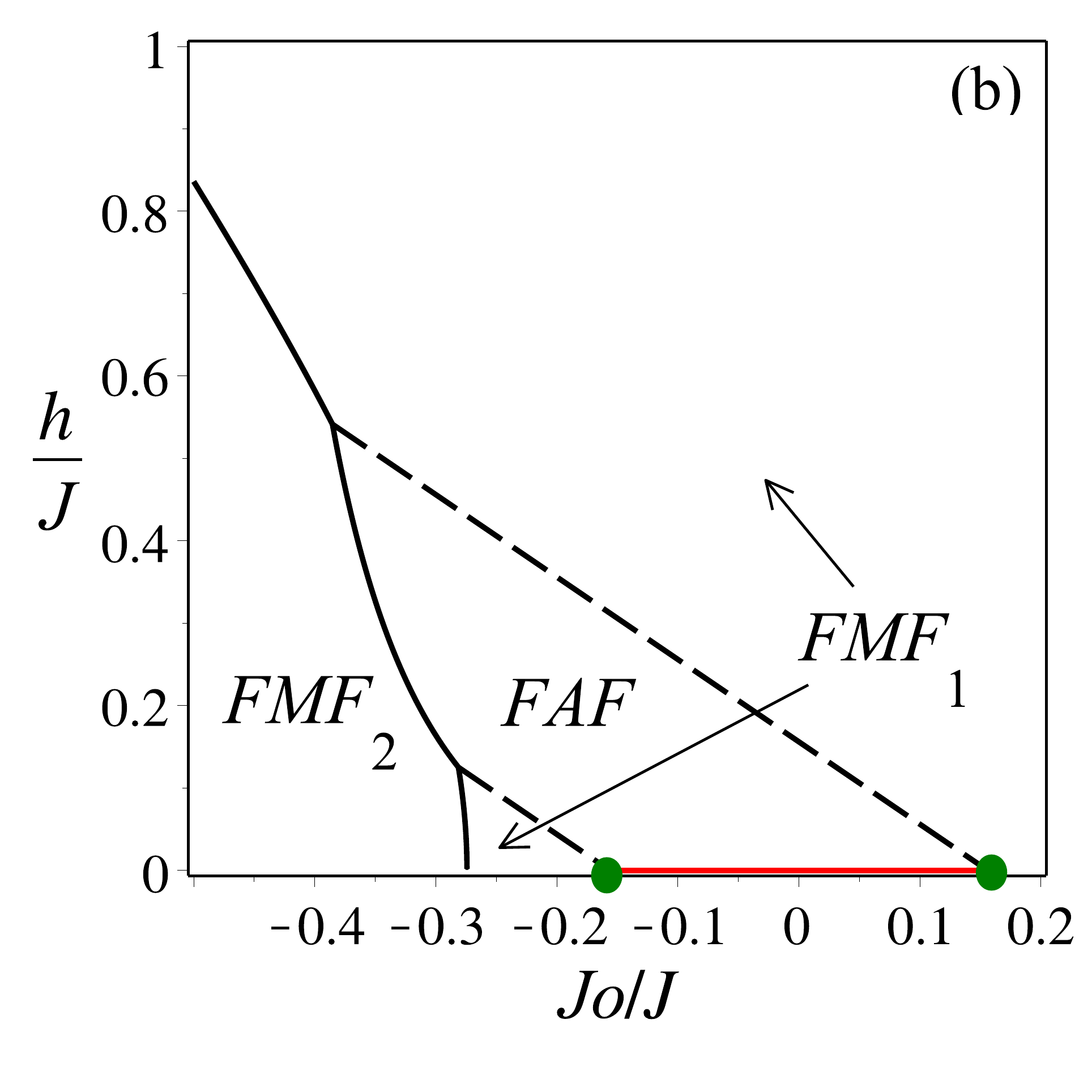}\protect\caption{\label{fig:Ph-Mg-0}Zero temperature phase diagram for the limiting
case Ising-XY ($J_{z}/J=0$ ) and in absence of magnetic field. (a)
Phase diagram $J_{0}/J$ against $\gamma$, for a fixed value $h/J=0$.
(b) For $J_{0}/J$ versus $h/J$ and assuming $\gamma=0.95$. }
\end{figure}

Here, we start our discussion regarding to the phase diagram at zero
temperature, first we assume the chain is in the absence of magnetic
field. Thus, we illustrate the phase diagram in figure \ref{fig:Ph-Mg-0}(a),
as a function of $J_{0}/J$ against $\gamma$ assuming a fixed value
$J_{z}/J=0$ and $h/J=0$. Where, we show two phases one Ising spin
ferromagnetic with Heisenberg spin modulated ferromagnetic simply
denoted by modulated ferromagnetic ($FMF$) phase, and the Ising spin
frustrated region ($IFR$).

The explicit representations of $FMF$ states are expressed below
\begin{alignat}{1}
|FMF_{1}\rangle= & \overset{N}{\underset{i=1}{\prod}}|\varphi_{4}\rangle_{i}\otimes|+\rangle_{i},\label{eq:FMF1}\\
|FMF_{2}\rangle= & \overset{N}{\underset{i=1}{\prod}}|\varphi_{4}\rangle_{i}\otimes|-\rangle_{i},\label{eq:FMF2}
\end{alignat}
and the ground state energy are given by
\begin{alignat}{1}
E_{FMF_{1}}= & -\frac{h}{2}-\sqrt{\left(h+J_{0}\right)^{2}+\tfrac{1}{4}J^{2}\gamma^{2}},\\
E_{FMF_{2}}= & \frac{h}{2}-\sqrt{\left(h-J_{0}\right)^{2}+\tfrac{1}{4}J^{2}\gamma^{2}},
\end{alignat}
it is worth to notice that $FMF_{1}$ and $FMF_{2}$ are degenerated
at null magnetic field, what we denote simply by $FMF$ region. At
first glance the states $FMF_{1}$ and $FMF_{2}$ seem equivalents
under total spin inversion; however, the state $|\varphi_{4}\rangle$
under spin inversion leads to a different state. Therefore, the states
$FMF_{1}$ and $FMF_{2}$ are no longer equivalents. 

Nevertheless, with no magnetic field, the model illustrates a frustrated
region, which is represented as
\begin{equation}
|IFR\rangle=\overset{N}{\underset{i=1}{\prod}}|\varphi_{2}\rangle_{i}\otimes|S\rangle_{i},\label{IFRv}
\end{equation}
with ground state energy given by $E_{IFR}=-\frac{|J|}{2}.$

Observe, that $IFR$ phase has a residual entropy $\mathcal{S}_{0}=\kappa_{B}\ln(2)$, the term 2 becomes from two possible orientation of  $S$ in  eq.\eqref{IFRv}.
The green curve contouring the IFR region given by $\gamma=\sqrt{1-4J_{0}^{2}}$
has a residual entropy $\mathcal{S}_{0}=\kappa_{B}\ln(3)$, while
the blue point corresponds to a residual entropy $\mathcal{S}_{0}=\kappa_{B}\ln(4)$.

A similar phase diagram at zero temperature is displayed in figure
\ref{fig:Ph-Mg-0}(b) as a function of $J_{0}/J$ and $h/J$ for a
fixed value $\gamma=0.95$ and $J_{z}/J=0$, the thick red line has
a residual entropy $\mathcal{S}_{0}=\kappa_{B}\ln(2)$, and the green
(gray) circle corresponds to a frustrated point with residual entropy
given by $\mathcal{S}_{0}=\kappa_{B}\ln(3)$, while the dashed line
curve given by $h/J=\pm\frac{\sqrt{39}}{40}-\frac{J_{0}}{J}$ (upper
lower respectively) has a residual entropy $\mathcal{S}_{0}=\kappa_{B}\ln(2)$.
\begin{figure}
\centering{}\includegraphics[scale=0.22]{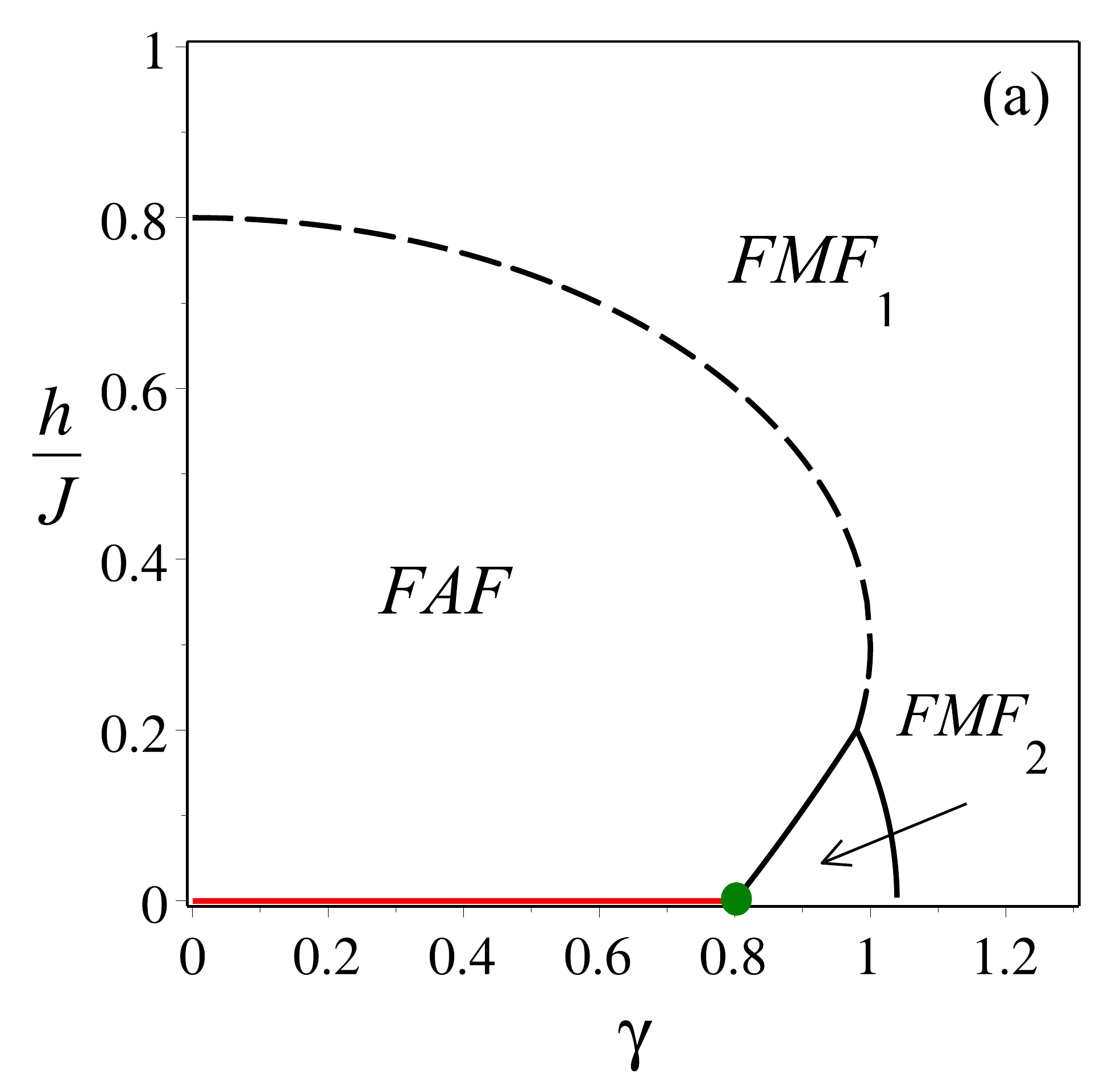}\includegraphics[scale=0.22]{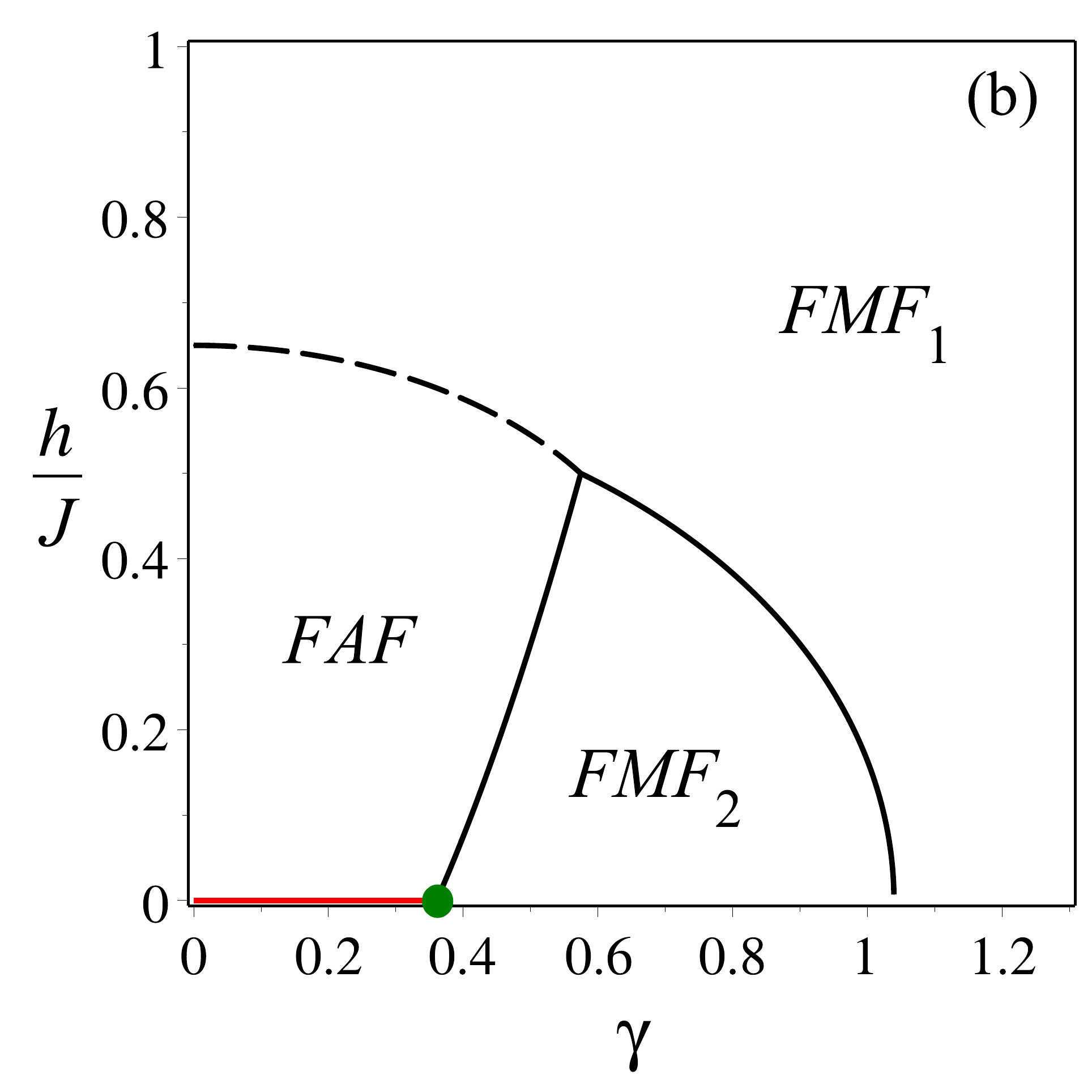}\protect\caption{\label{fig:PD t=00003D0}Phase diagram at zero temperature for the
Ising-$XY$ diamond chain in units of $J$, for $\gamma$ versus $h/J$,
assuming $J_{0}/J=-0.3$. (a) $J_{z}/J=0$. b) $J_{z}/J=0.3$.}
\end{figure}

In figure \ref{fig:PD t=00003D0}(a) is displayed the phase diagram
at zero temperature $\gamma$ against $h/J$ for a fixed value $J_{0}/J=-0.3$
and $J_{z}/J=0$, the thick red line corresponds to Ising frustrated
($IFR$) region in a zero magnetic field, while the dashed line corresponds
to a non-zero Heisenberg frustrated ($HFR$) phase. Furthermore, is
displayed three phase: one $FAF$ phase,$FMF_{1}$, and another
$FMF_{2}$ phase, given by
states \eqref{eq:FMF1} and \eqref{eq:FMF2}. Similarly, phase diagram
at zero temperature is displayed in figure \ref{fig:PD t=00003D0}(b)
as a function of $\gamma$ and $h/J$ for fixed value $J_{0}/J=-0.3$
and $J_{z}/J=0.3$, the thick red line and dashed line and the green (gray) circle have the same
meaning as for the case (a).

An explicit representation of $FAF$ state is expressed below
\begin{alignat}{1}
|FAF\rangle= & \overset{N}{\underset{i=1}{\prod}}|\varphi_{2}\rangle_{i}\otimes|+\rangle_{i},
\end{alignat}
and its ground state energy is given by
$
E_{FAF}=  -\frac{J+h}{2}.
$

Particularly, when $\gamma=\sqrt{1-4(J_{0}+h)^{2}}$, we obtain frustration
state with magnetic field, given by the state
\begin{equation}
|HFR\rangle=\overset{N}{\underset{i=1}{\prod}}|\varphi_{2(4)}\rangle_{i}\otimes|+\rangle_{i},
\end{equation}
with its corresponding ground state energy given by
$
E_{HFR}=-\frac{|J|+h}{2}.
$

It is worth to mention that $HFR$ phase (dashed line in figures \ref{fig:Ph-Mg-0}
and \ref{fig:PD t=00003D0}) has also a residual entropy $\mathcal{S}_{0}=\kappa_{B}\ln(2)$. At low temperatures residual entropies have common features with thermal entanglement. Comparing the fig. \ref{fig:PD t=00003D0}(b) and the fig. \ref{fig:Den-entng}(a) in next section one can observe that fact.

One can also easily verify the $ab$-dimer are entangled in all region
illustrated in the phase diagram, only limiting cases could lead to 
disentangled regions such as the discussed in Ising-XXZ chain\cite{spra}. 

\section{Thermodynamics}

The Ising-XYZ diamond chain can be solved exactly using the decoration
transformation\cite{Syozi,Fisher,phys-A-09,strecka pla} together
with the usual transfer matrix technique \cite{baxter-book}. So,
let us start considering the partition function as follow
\begin{equation}
\mathbb{\mathcal{Z}}={\rm tr}\left({\rm e}^{-\beta H}\right),
\end{equation}
where $\beta=1/\kappa_{B}T$, with $\kappa_{B}$ being the Boltzmann
constant and $T$ is the absolute temperature, and the Hamiltonian
$H$ is given by eq.\eqref{eq:Hamt}. The transfer matrix of the model
can be expressed by 
\begin{equation}
T=\left[\begin{array}{cc}
w(2) & w(0)\\
w(0) & w(-2)
\end{array}\right],
\end{equation}
where the Boltzmann factors is expressed by
\begin{equation}
w(\mu)=2{\rm e}^{\frac{\beta\mu h}{2}}\left[{\rm e}^{-\frac{\beta J_{z}}{4}}\cosh\left(\tfrac{\beta J}{2}\right)+{\rm e}^{\frac{\beta J_{z}}{4}}\cosh\left(\beta\Delta(\mu)\right)\right],
\end{equation}
and transfer matrix eigenvalues, become
\begin{equation}
\lambda_{\pm}=w(2)+w(-2)\pm\sqrt{\left(w(2)-w(-2)\right)^{2}+4w(0)^{2}}.
\end{equation}

To study the thermodynamic properties we will use the exact free energy
per unit cell in thermodynamic limit 
\[
f=-\frac{1}{\beta}\underset{N\rightarrow+\infty}{\lim}\frac{\ln\mathcal{Z}}{N}=-\frac{1}{\beta}\ln\lambda_{+}.
\]

Next, let us calculate the thermal entanglement behavior between $ab$-dimer
of our model investigated.

\section{Pairwise Thermal Entanglement}

Now let us start our discussion regarding the quantum entanglement
of the Ising-XYZ diamond chain, recalling that a two-qubits with XYZ
coupling was discussed in reference \cite{G. Rigolin}. As a measure
of entanglement for two arbitrary mixed states of dimers, we use the
quantity called concurrence\cite{wootters}, which is defined in terms
of reduced density matrix $\rho$ of two mixed states
\begin{equation}
\mathcal{C}(\rho)=\max\{{0,2\Lambda_{{\rm max}}-\text{{tr}}\sqrt{R}}\},
\end{equation}
assuming
$
R=\rho\sigma^{y}\otimes\sigma^{y}\rho^{*}\sigma^{y}\otimes\sigma^{y},
$
where $\Lambda_{{\rm max}}$ is the largest eigenvalue of the matrix
$\sqrt{R}$ and $\rho^{*}$ represent the complex conjugate of matrix
$\rho$, with $\sigma^{y}$ being the Pauli matrix.

Consequently the concurrence between $ab$-dimer becomes
\begin{equation}
\mathcal{C}=\max\{0,|\rho_{14}|-{\rho_{22}\rho_{33}},|\rho_{23}|-\sqrt{\rho_{11}\rho_{44}}\},\label{eq:Cnr-ab}
\end{equation}
where the $\rho_{ij}$ are the elements of the density matrix. For
the case of an infinite chain, the reduced density operator elements\cite{bukman}
could be expressed in terms of the correlation function between two
entangled particles\cite{amico},
\begin{alignat}{1}
\rho_{11}= & \tfrac{1}{4}+\langle\sigma_{a}^{z}\sigma_{b}^{z}\rangle+\langle\sigma_{a}^{z}\rangle,\\
\rho_{22}=\rho_{33}= & \tfrac{1}{4}-\langle\sigma_{a}^{z}\sigma_{b}^{z}\rangle,\\
\rho_{44}= & \tfrac{1}{4}+\langle\sigma_{a}^{z}\sigma_{b}^{z}\rangle-\langle\sigma_{a}^{z}\rangle,\\
\rho_{14}= & \langle\sigma_{a}^{x}\sigma_{b}^{x}\rangle-\langle\sigma_{a}^{y}\sigma_{b}^{y}\rangle,\\
\rho_{23}= & \langle\sigma_{a}^{x}\sigma_{b}^{x}\rangle+\langle\sigma_{a}^{y}\sigma_{b}^{y}\rangle.
\end{alignat}

It is worth to mention that the zero temperature entanglement for
$ab$-dimer is maximally entangled in $FAF$ region $\mathcal{C}=1$,
while in the region $FMF$ the concurrence depends on Hamiltonian
parameters, which is given by 
$
\mathcal{C}=\frac{|J\gamma|}{2\sqrt{\left(J_{0}+h\right)^{2}+\frac{1}{4}J^{2}\gamma^{2}}}.
$

However, at finite temperature, each expected values become temperature
dependent quantities which are expressed as
\begin{alignat}{1}
\langle\sigma_{a}^{x}\sigma_{b}^{x}\rangle= & {\rm e}^{\beta\frac{h}{2}}\tfrac{\tfrac{\Delta(1)}{2}{\rm e}^{-\beta\frac{2h+J_{z}}{4}}\sinh(\tfrac{\beta J}{2})+\tfrac{J\gamma}{4}{\rm e}^{\beta\frac{J_{z}}{4}}\sinh(\beta\Delta(1))}{\Delta(1)\lambda_{+}},\\
\langle\sigma_{a}^{y}\sigma_{b}^{y}\rangle= & {\rm e}^{\beta\frac{h}{2}}\tfrac{\tfrac{\Delta(1)}{2}{\rm e}^{-\beta\frac{J_{z}}{4}}\sinh(\tfrac{\beta J}{2})-\tfrac{J\gamma}{4}{\rm e}^{\beta\frac{J_{z}}{4}}\sinh(\beta\Delta(1))}{\Delta(1)\lambda_{+}},\\
\langle\sigma_{a}^{z}\sigma_{b}^{z}\rangle= & {\rm e}^{\beta\frac{h}{2}}\tfrac{{\rm e}^{\beta\frac{J_{z}}{4}}\cosh(\tfrac{\beta J}{2})-{\rm e}^{-\beta\frac{J_{z}}{4}}\cosh(\beta\Delta(1))}{2\lambda_{+}},\\
\langle\sigma_{a}^{z}\rangle= & {\rm e}^{\beta\frac{2h+J_{z}}{4}}\sinh(\beta\Delta(1))\frac{J_{0}+h}{\Delta(1)\lambda_{+}},
\end{alignat}

Alternatively, we can obtain also equivalent result using the approach
described in reference \cite{spra}.
\begin{figure}
\includegraphics[scale=0.2]{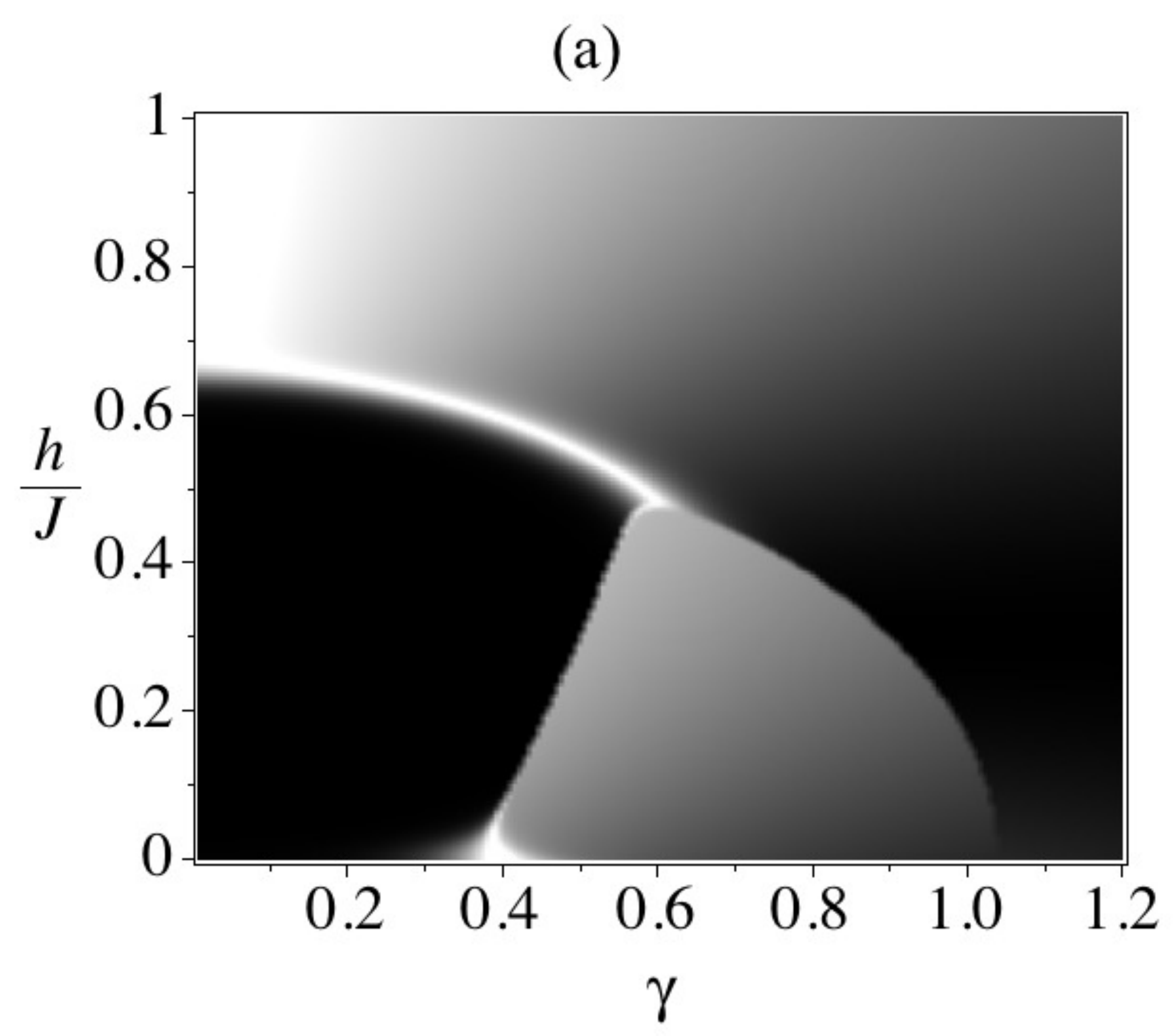}\includegraphics[scale=0.2]{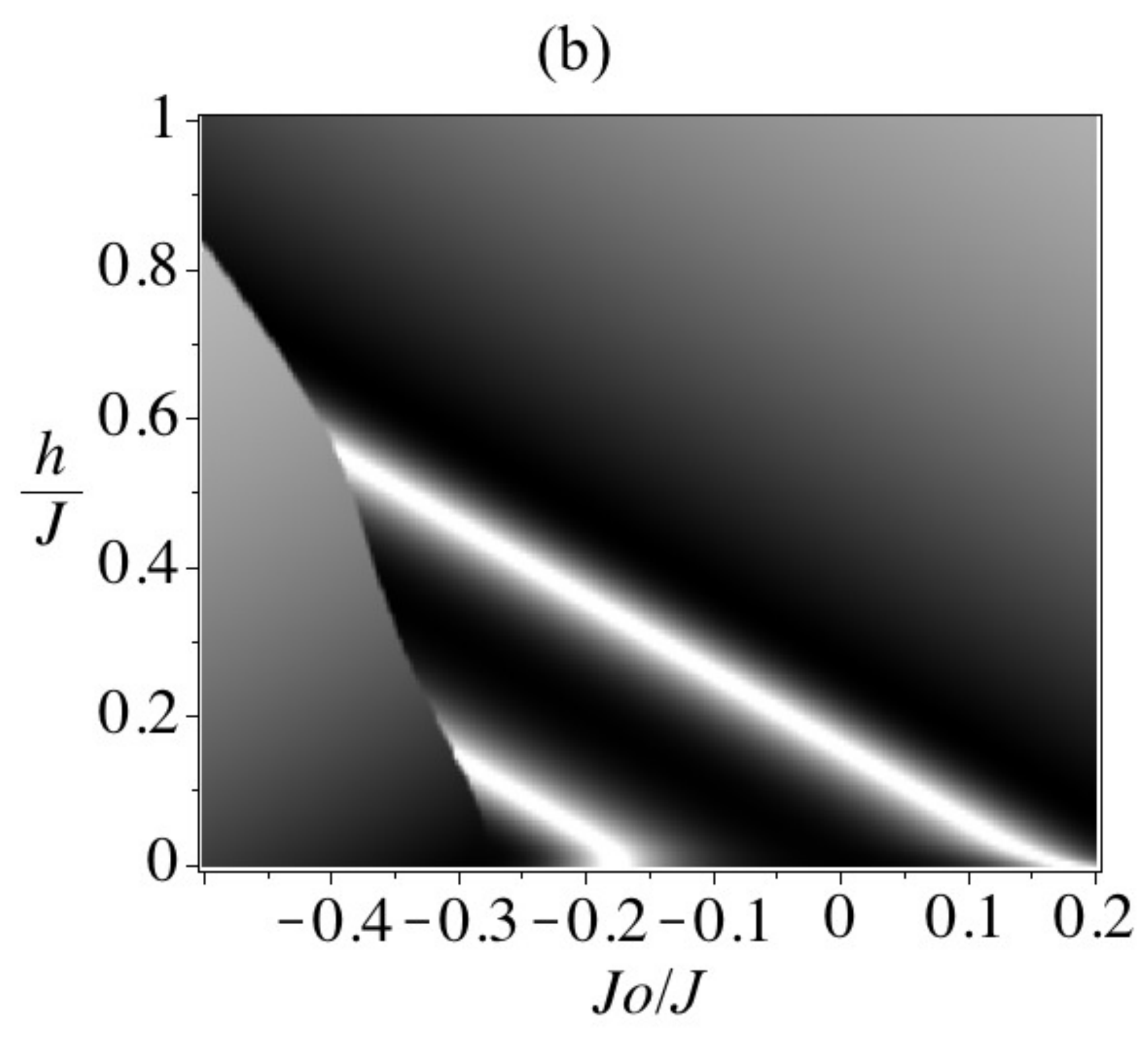}
\protect\caption{\label{fig:Den-entng}Density-plot thermal entanglement in the low
temperature limit $T/J=0.01$. (a) Concurrence as a function of $\gamma$
against $h/J$ assuming $J_{z}/J=0.3$. (b) Concurrence $J_{0}/J$
versus $h/J$ for the Isin-XY diamond chain ($J_{z}/J=0$ and $\gamma=0.95$)$ $.}
\end{figure}

In figure \ref{fig:Den-entng}(a) is illustrated a density-plot concurrence
in the low temperature limit $T/J=0.01$ with fixed value $J_{0}/J=-0.3$
and $J_{z}/J=0.3$, as a function of $\gamma$ and $h/J$. The darkest
region corresponds to a higher concurrence ($\mathcal{C}=1$), while
white region represents untangled region ($\mathcal{C}=0$), this
plot follow the pattern of the phase diagram displayed in figure \ref{fig:PD t=00003D0}(a).
Whereas, in figure \ref{fig:Den-entng}(b) is illustrated a density-plot
concurrence in the low temperature limit $T/J=0.01$ with fixed value
$J_{z}/J=0$ and $\gamma=0.95$, as a function of $J_{0}/J$ and $h/J$.
This plot also follows the pattern of the phase diagram illustrated
in figure \ref{fig:Ph-Mg-0}(b). It is worth to mention that the disentangled
region has similar behaviour to those found in reference \cite{G. Rigolin} for null magnetic field and  with external magnetic field in reference \cite{Zhou-Song}, both for the case of two qubits with XYZ coupling.
\begin{figure}
\includegraphics[scale=0.28]{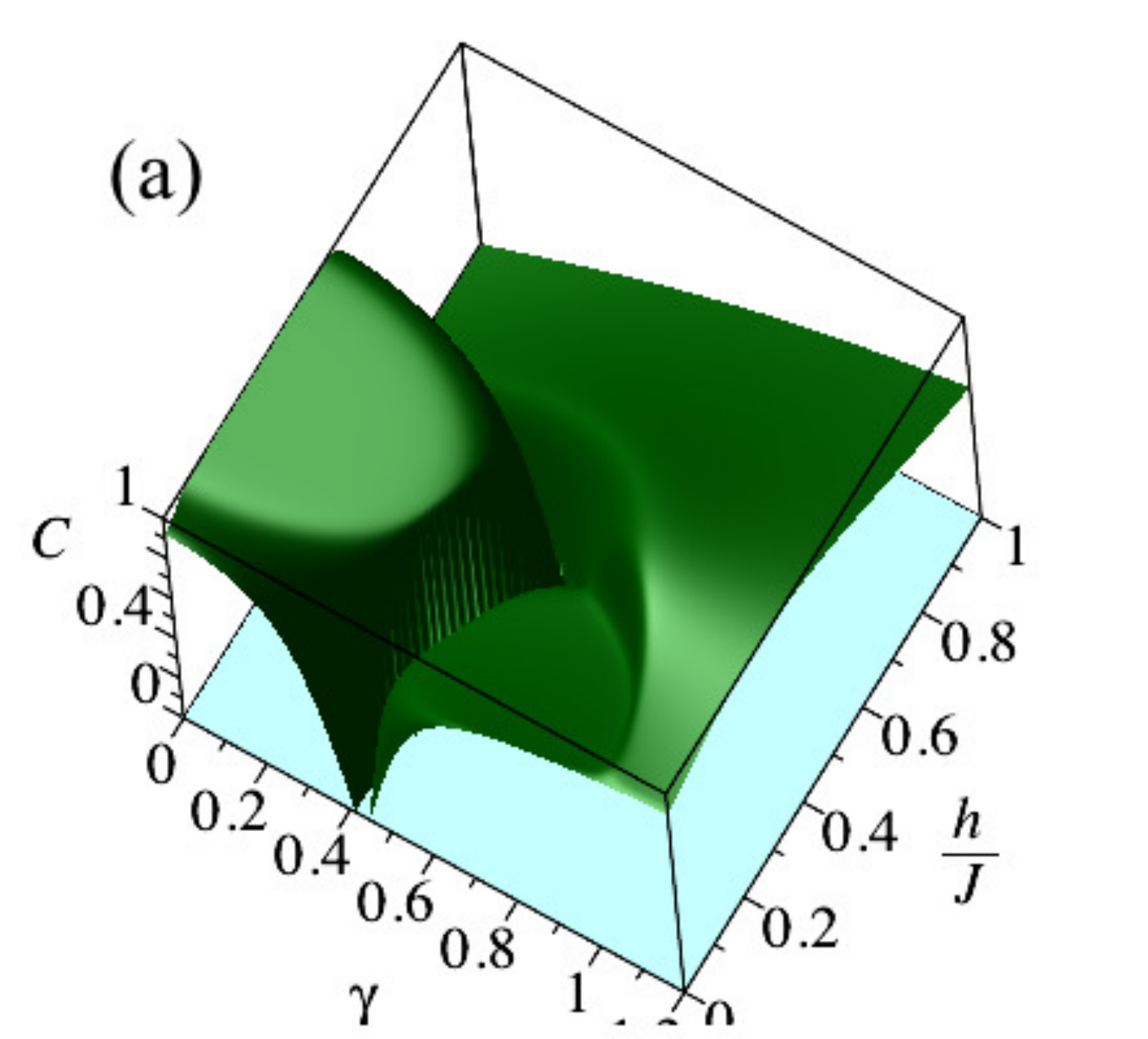}\includegraphics[scale=0.28]{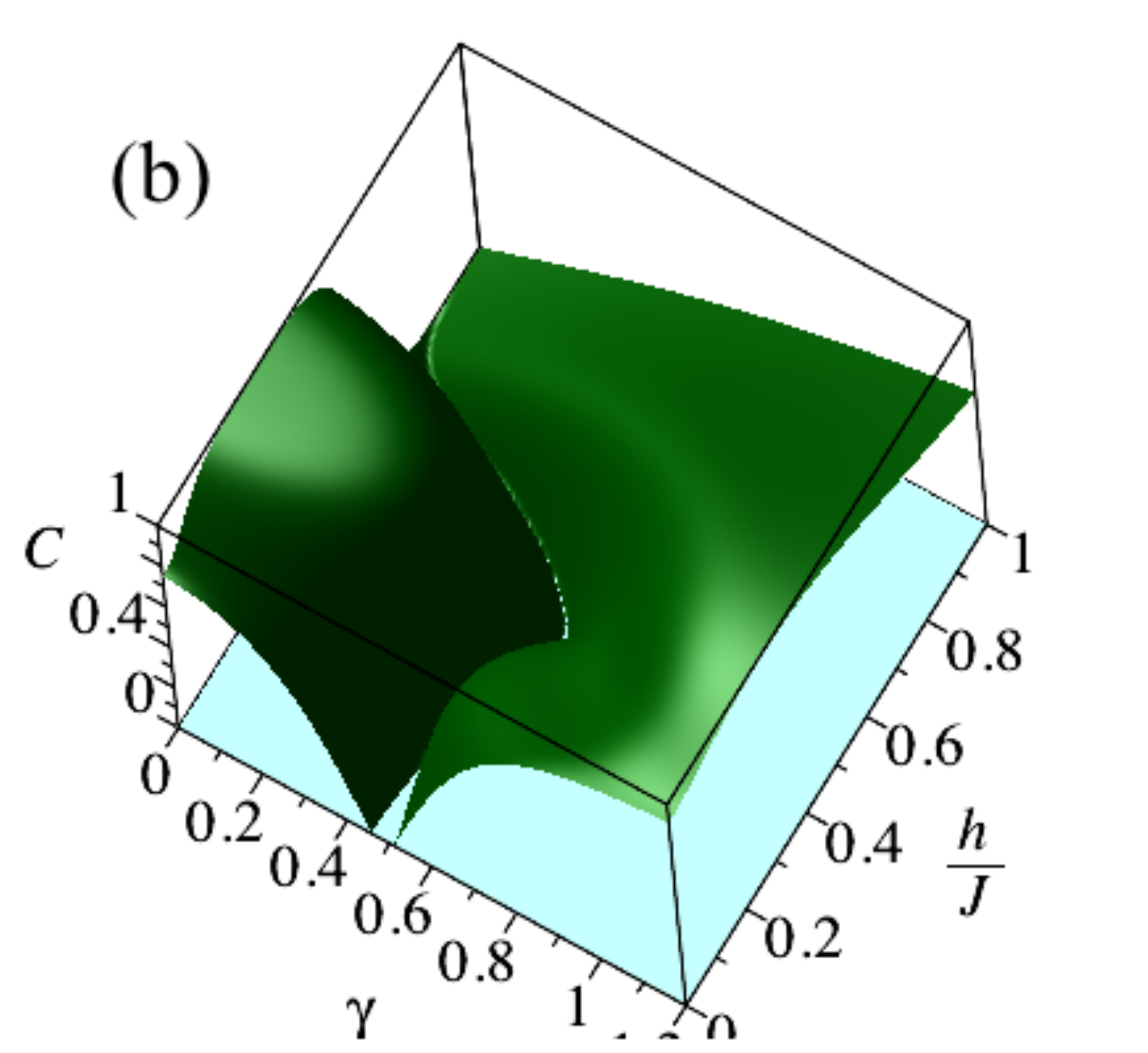}
\includegraphics[scale=0.28]{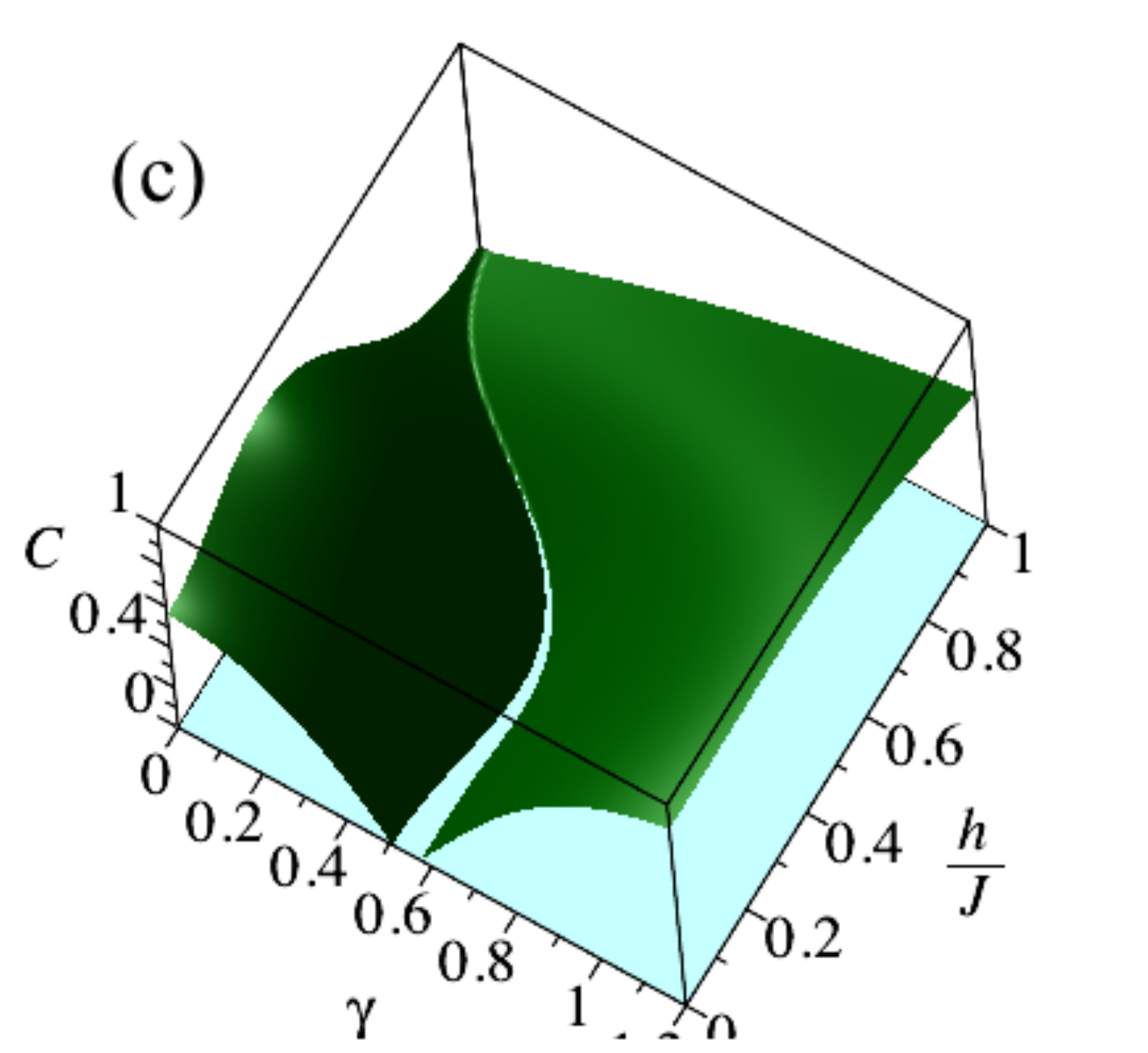}\includegraphics[scale=0.28]{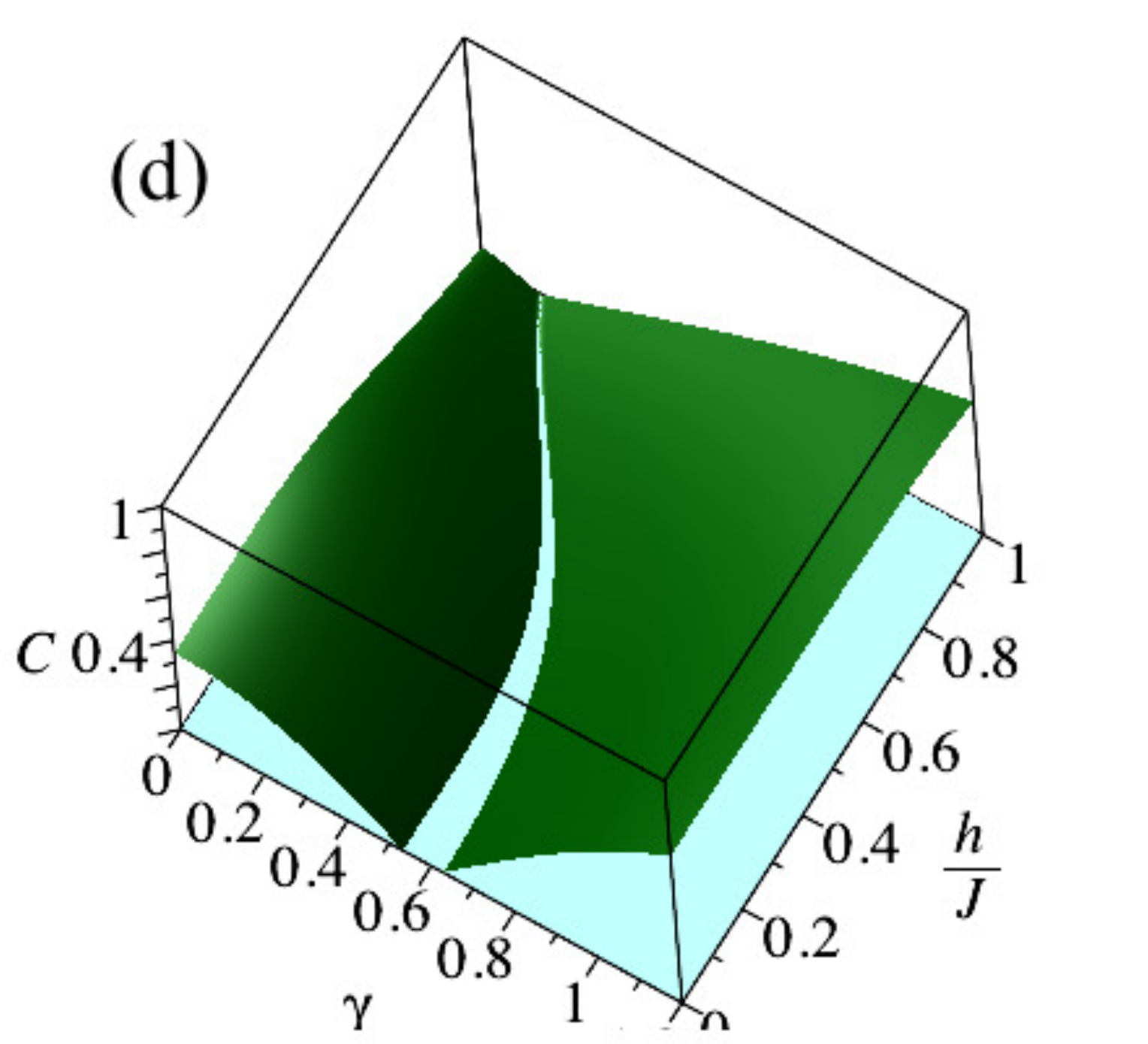}
\protect\caption{\label{fig:entg-detr}(Color Online) Thermal entanglement deterioration
as a function of $\gamma$ and $h/J$ when temperature increases and
fixed $J_{0}/J=-0.3$ and $J_{z}/J=0.3$. (a) $T/J=0.02$. (b) $T/J=0.04$.
(c) $T/J=0.08$. (d) $T/J=0.16$.}
\end{figure}

For the purpose to give a more detailed behavior of the concurrence
as a function of $\gamma$ and $h/J$. In figure \ref{fig:entg-detr}
is illustrated the concurrence as a sequence of higher temperature
than figure \ref{fig:Den-entng}(a), assuming the concurrence for
same fixed parameters $J_{0}/J=-0.3$ and $J_{z}/J=0.3$. In fig.\ref{fig:entg-detr}(a)
the thermal entanglement is illustrated at $T/J=0.02$, basically
still remains the entangled region (figure \ref{fig:Den-entng}(a))
and displaying the change of disentangled region. In fig.\ref{fig:entg-detr}(b)
is illustrated for $T/J=0.04$, showing how the entangled region \textquotedbl{}deteriorate\textquotedbl{}
and how the disentangled region is modified. Similarly in fig.\ref{fig:entg-detr}(c)
is illustrated for $T/J=0.08$, where entangled region becoming increasingly
deteriorate. Lastly in fig.\ref{fig:entg-detr}(d) is shown for $T/J=0.16$,
at this temperature the entangled region was highly modified compared
to that figure \ref{fig:Den-entng}(a), showing only a nearly straight
valley. In summary, we illustrate how the thermal entanglement \textquotedbl{}deteriorate\textquotedbl{}
(vanishes) as far as the temperature increases.

\subsection{Threshold temperature}

We follow a similar definition of the critical temperature for the threshold temperature
$T_{th}$, such as disorder-order-disorder (or inverse) sequences of transition
(reentrant phenomenon) have been observed in many systems. The reentrant
phase transition has been observed for the first time in polymer gels
\cite{katayama}, frustrated spin-gas model \cite{berker}, Rochelle
salts \cite{levitskii}, and a predator - prey model \cite{Han}.
Quantum reentrant phase transitions (disentangled (D) - entangled
(E) - disentangled (D) - entangled (E) - disentangled (D) regions)
of the concurrence versus magnetic field has been obtains noticed
in Lipkin-Meshkov-Glick model \cite{Morrison}.

\begin{figure}
\includegraphics[scale=0.28]{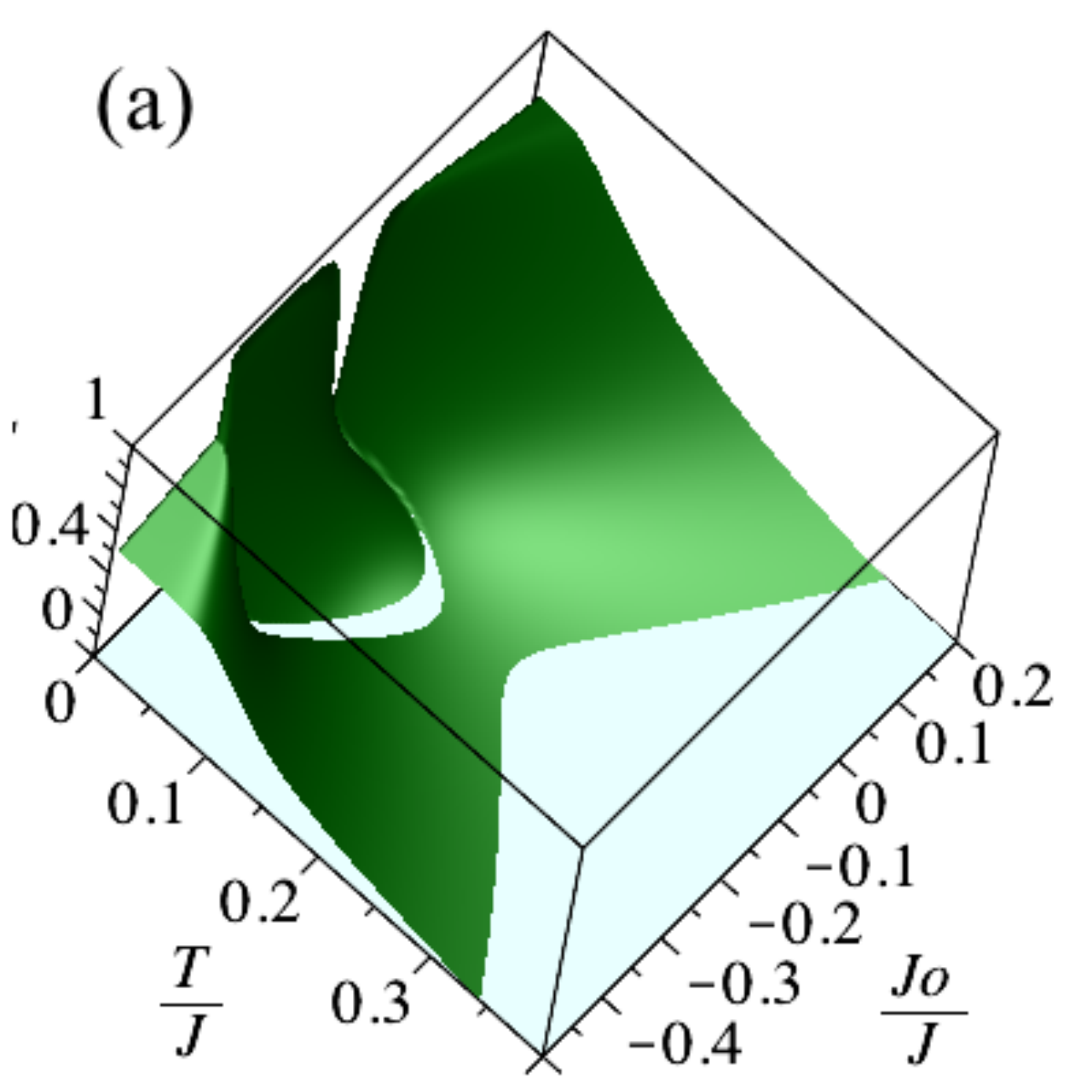}\includegraphics[scale=0.28]{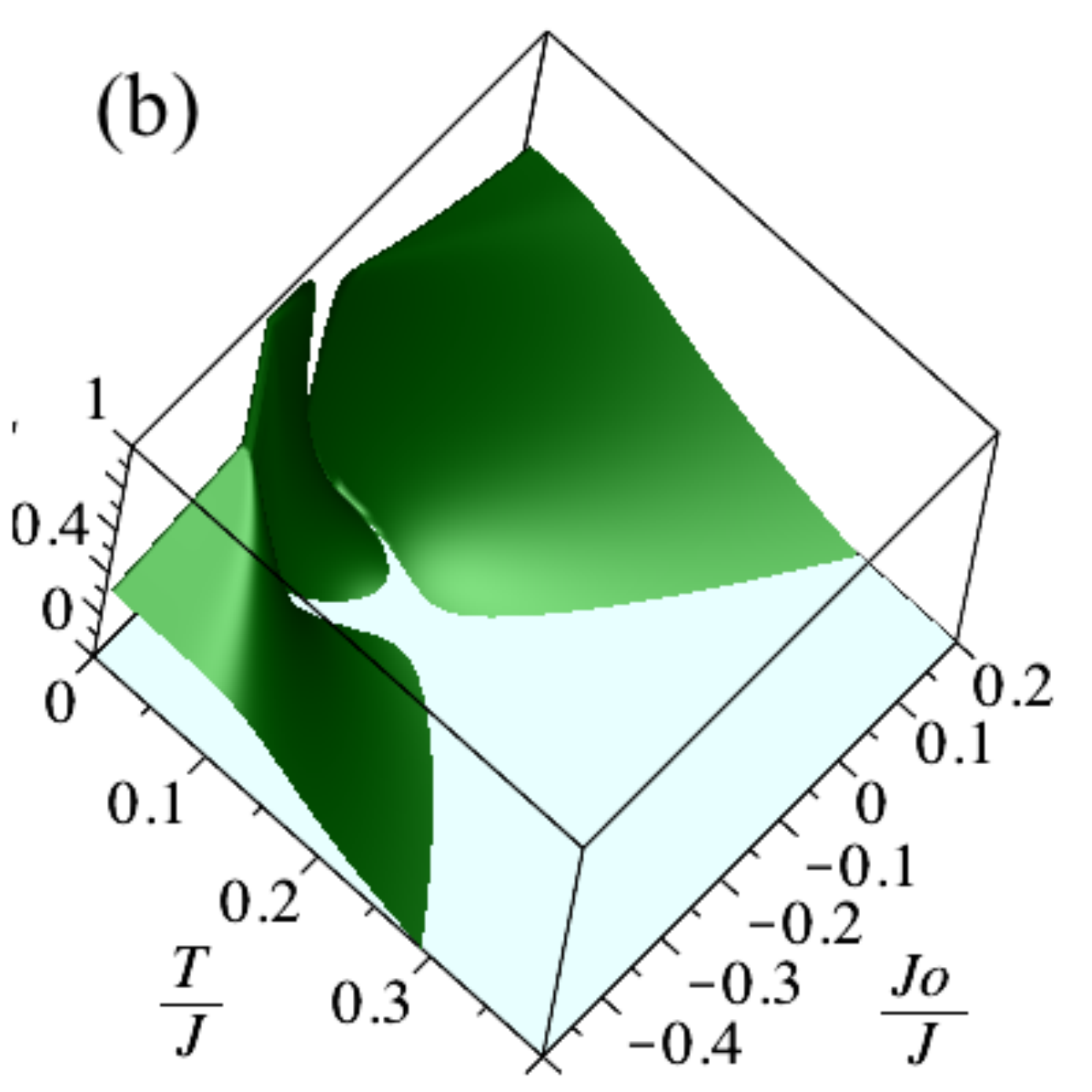}

\protect\caption{\label{fig:ent-tmp}(Color Online) Thermal entanglement deterioration
when temperature increases. (a) For $J_{z}/J=0$, $\gamma=0.95$ and
$h/J=0.27$. (b) For $J_{z}/J=0.3$, $\gamma=0.6$ and $h/J=0.35$.
 }
\end{figure}

In figure \ref{fig:ent-tmp}(a) is illustrated the concurrence as
a function of $T/J$ and $J_{0}/J$ assuming a fixed value $J_{z}/J=0$,
$\gamma=0.95$ and $h/J=0.27$, here we can observe how the concurrence
vanishes and for a higher temperature arise a thermal entanglement.
Thereafter, the entanglement for even a higher temperature disappears
altogether. In figure \ref{fig:ent-tmp}(b) is depicted the concurrence
as a function of $T/J$ and $J_{0}/J$ assuming a fixed value $J_{z}/J=0.3$,
$\gamma=0.6$ $h/J=0.35$, we can also observe, how the concurrence
vanishes for higher temperature and for a bit higher temperature emerges
a small thermal entanglement, hereafter the entanglement for higher
temperature disappears definitely. 

\begin{figure}
\includegraphics[width=2.8cm,height=2.8cm]{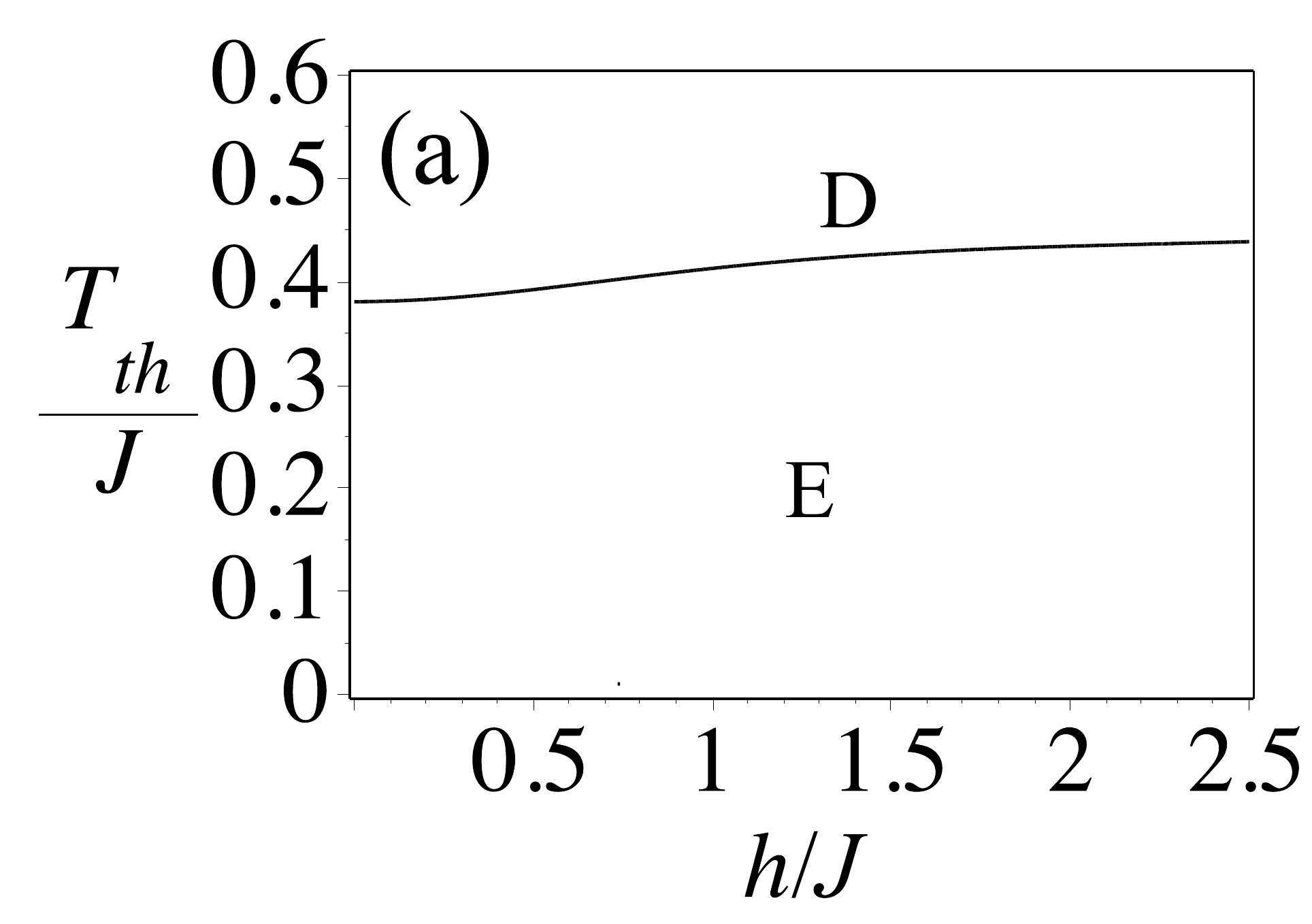}\includegraphics[width=2.8cm,height=2.8cm]{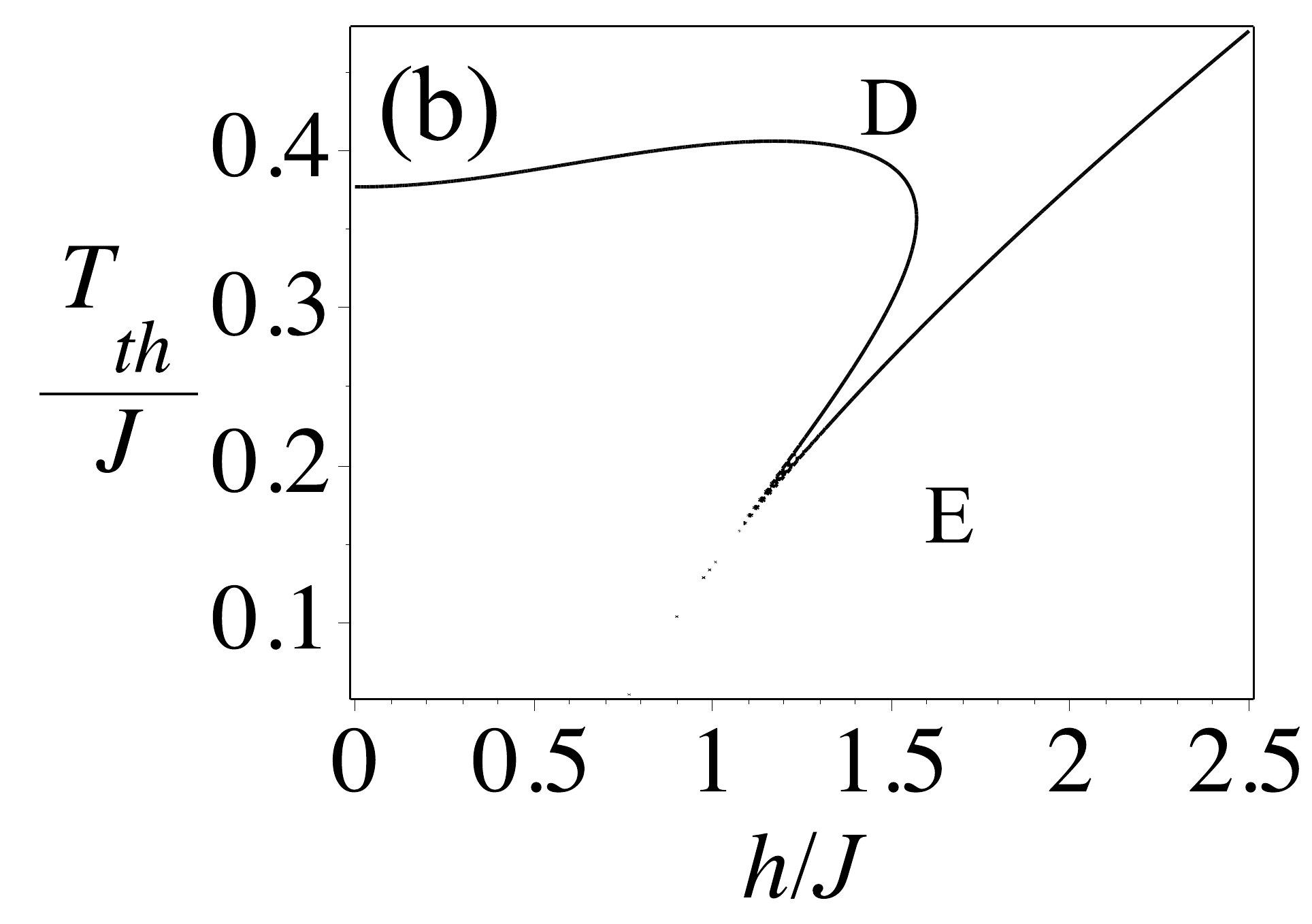}\includegraphics[width=2.8cm,height=2.8cm]{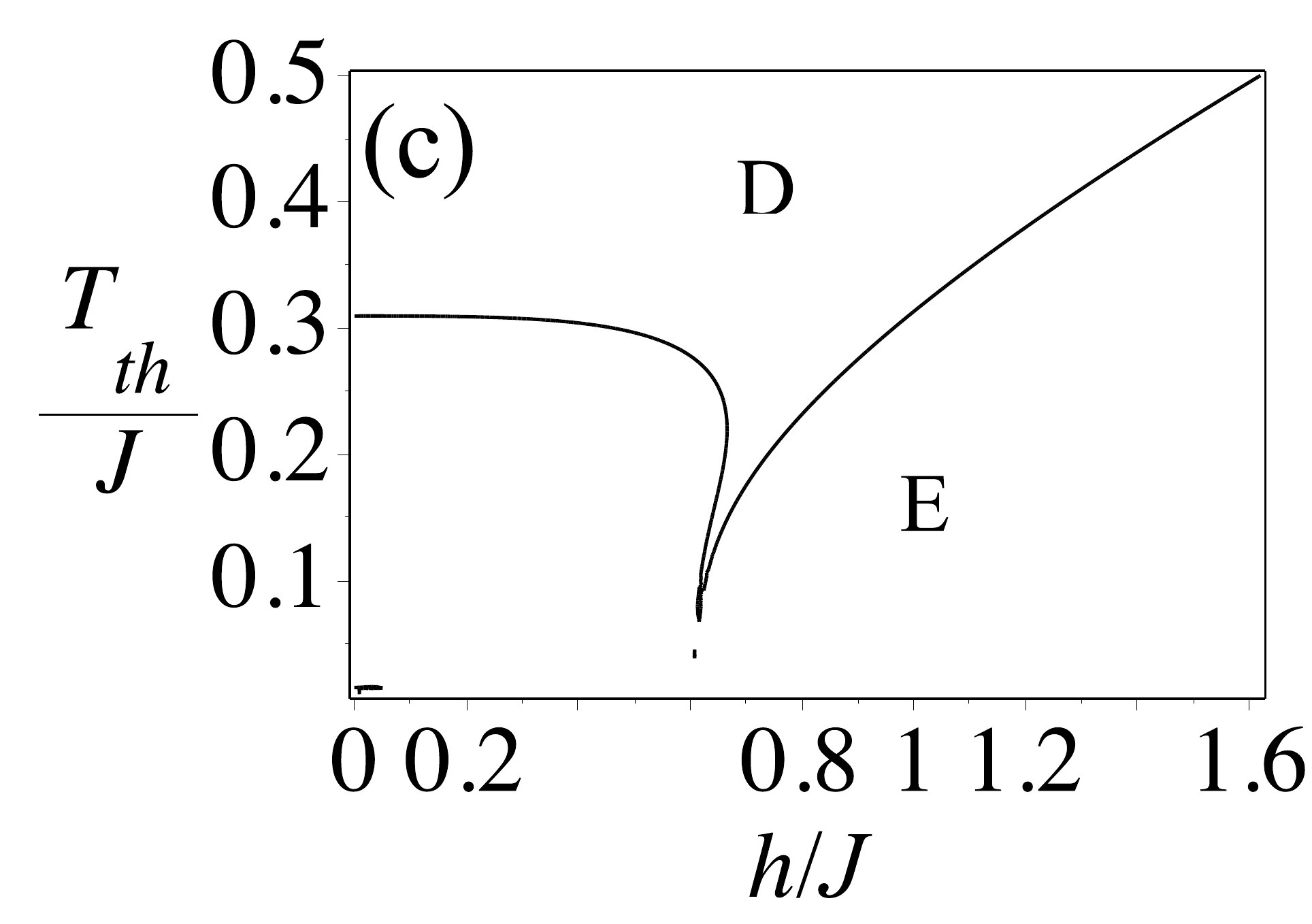}

\includegraphics[width=2.8cm,height=2.8cm]{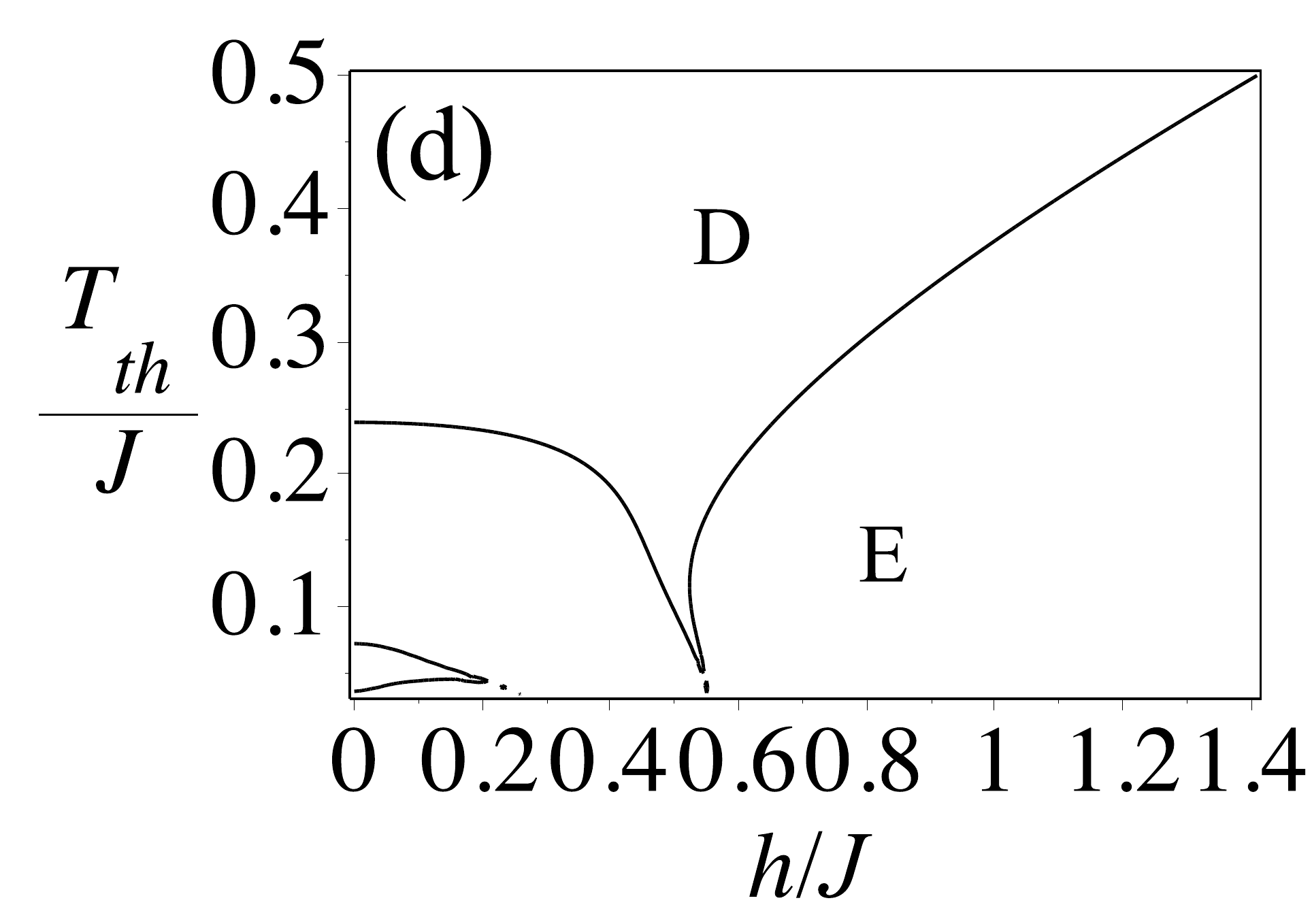}\includegraphics[width=2.8cm,height=2.8cm]{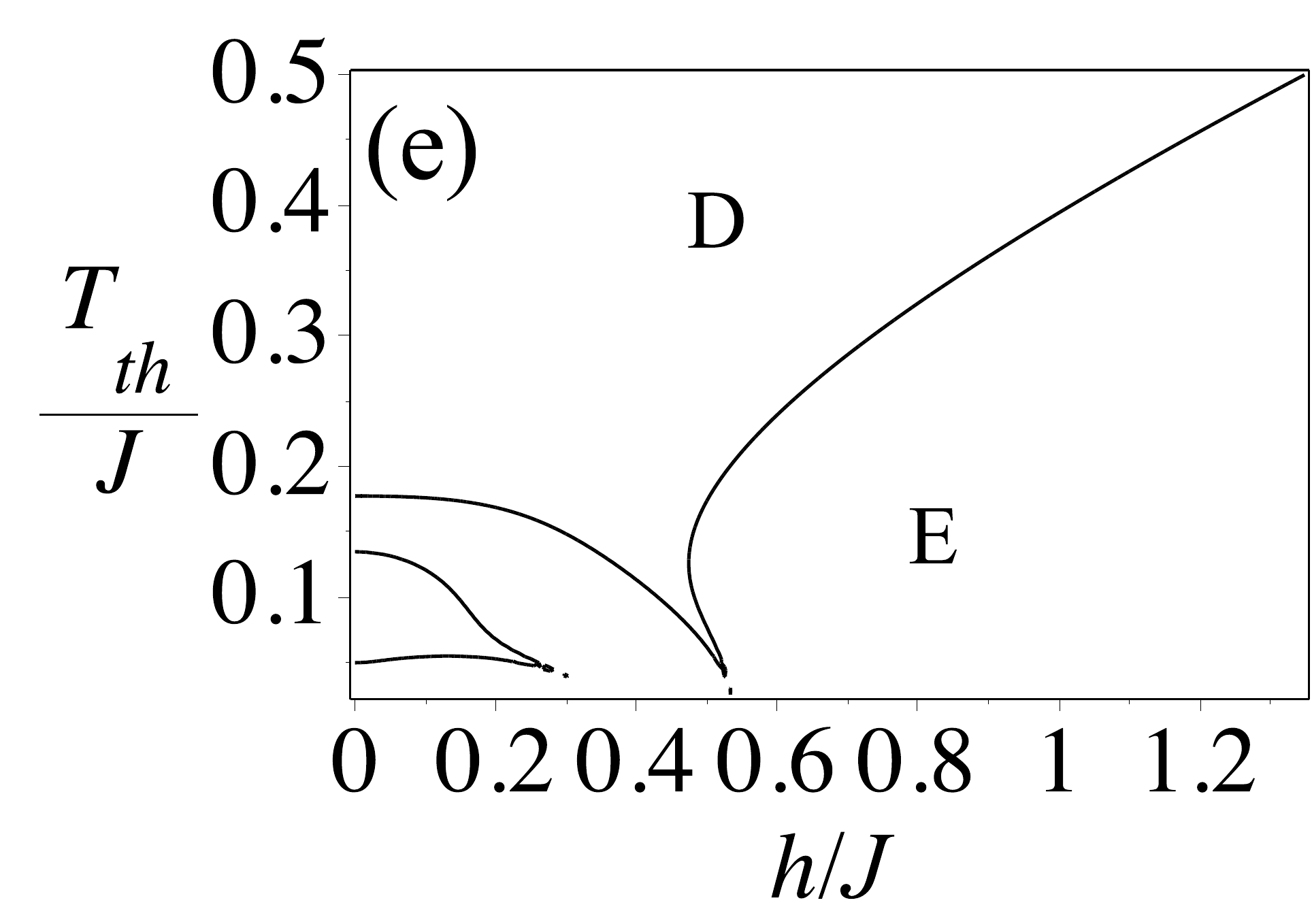}\includegraphics[width=2.8cm,height=2.8cm]{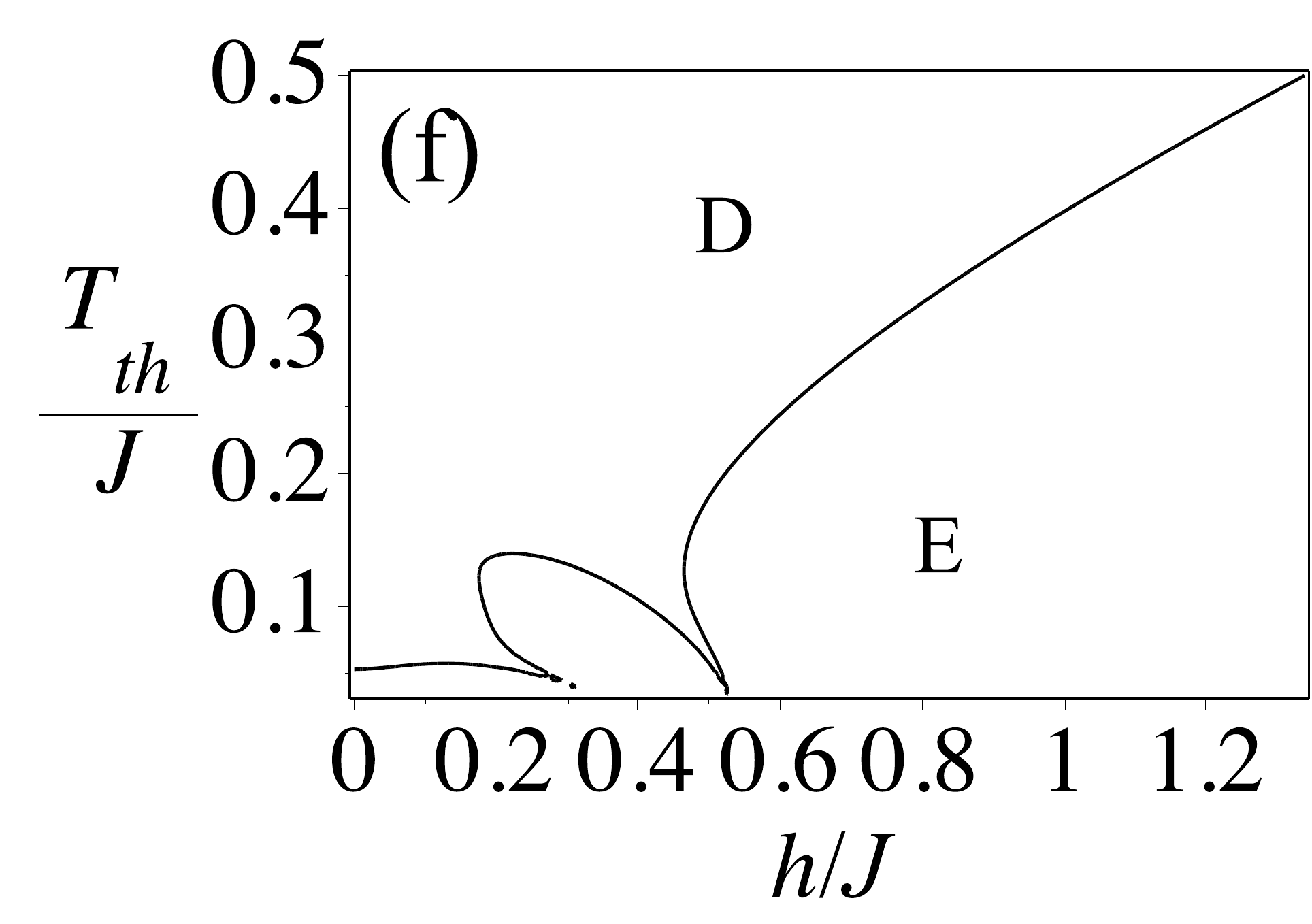}

\includegraphics[width=2.8cm,height=2.8cm]{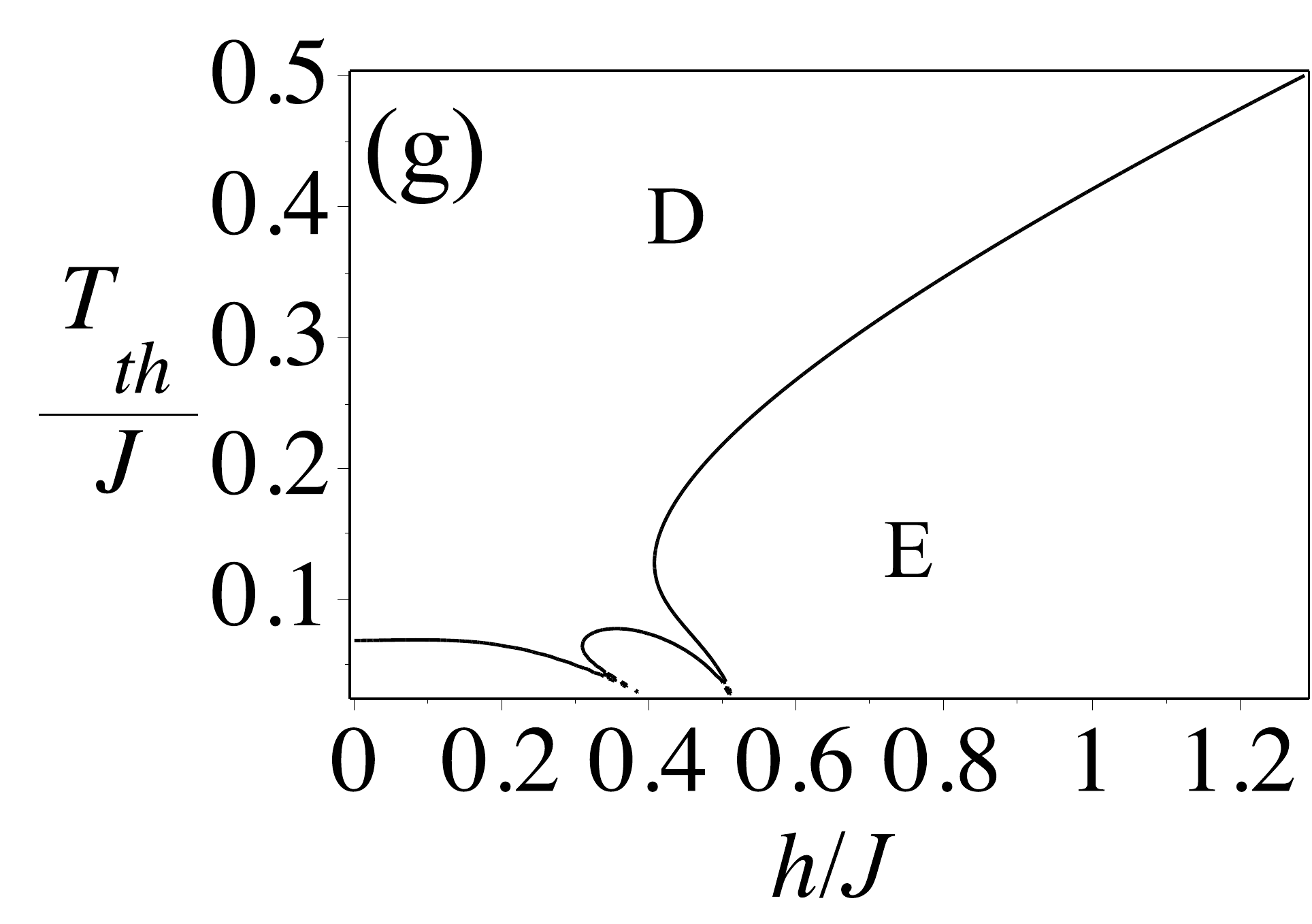}\includegraphics[width=2.8cm,height=2.8cm]{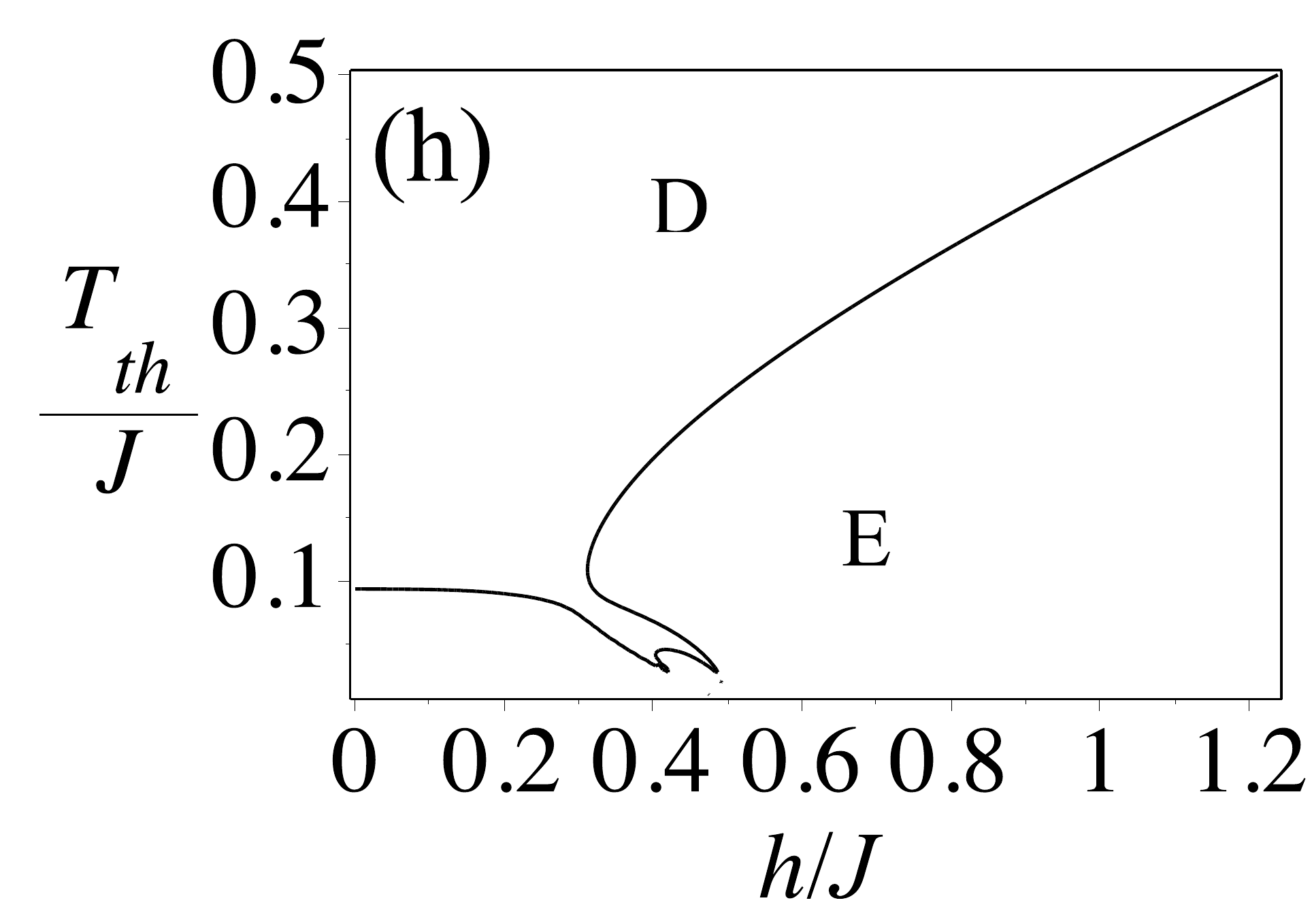}\includegraphics[width=2.8cm,height=2.8cm]{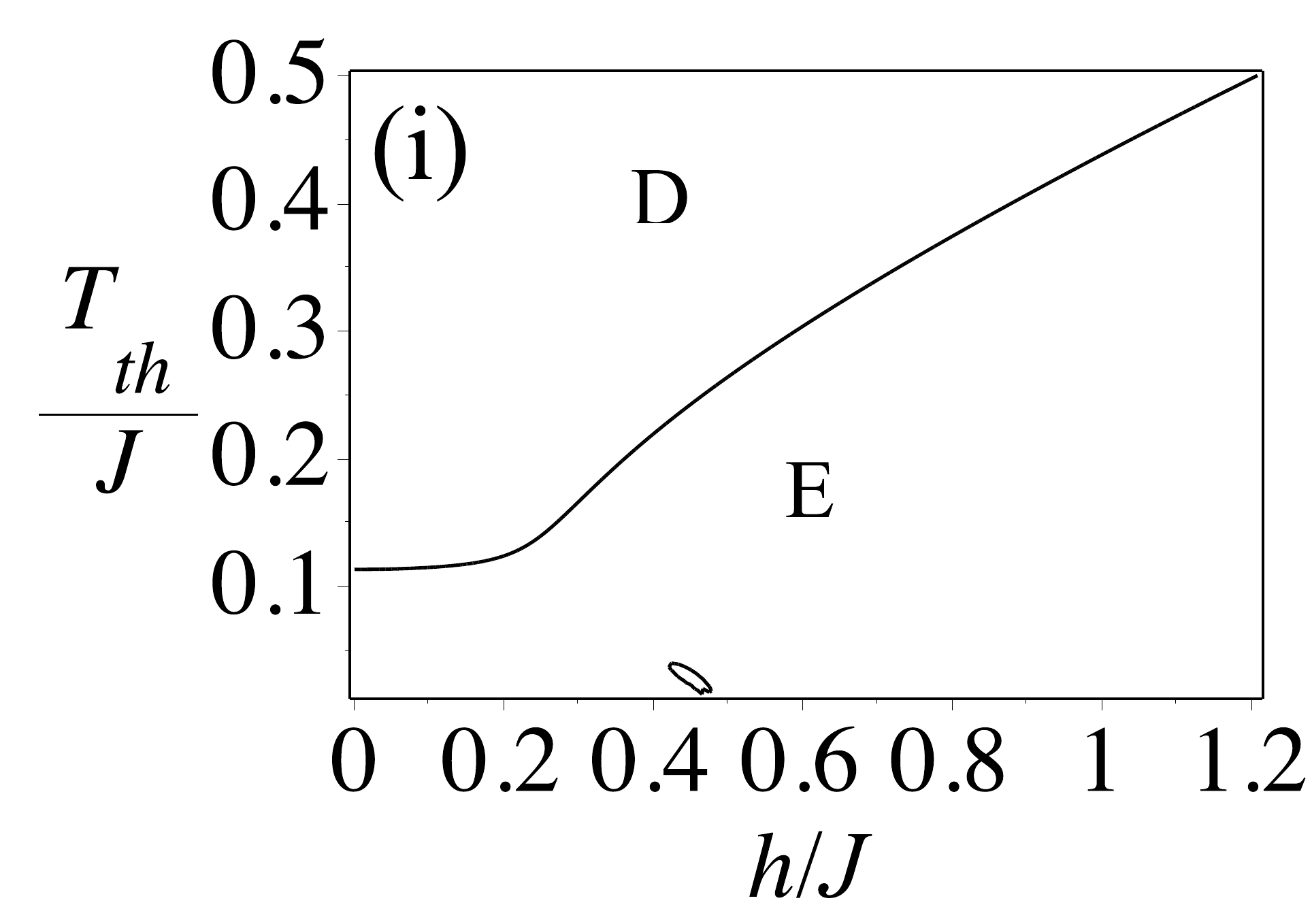}\protect\caption{\label{fig:Threshold}Threshold temperature limiting entangled (E)
and disentangled (D) region as a function of magnetic field, assuming
fixed parameters values $J_{0}/J=-0.3$ and $J_{z}/J$=0.3, for several
values of XY-anisotropy $\gamma$. (a) $\gamma=0$, (b) $\gamma=0.1$,
(c) $\gamma=0.4$, (d) $\gamma=0.5$, (e) $\gamma=0.53$, (f) $\gamma=0.535$,
(g) $\gamma=0.56$, (h) $\gamma=0.585$, (i) $\gamma=0.6$. }
\end{figure}

 In figure \ref{fig:Threshold}
is illustrated the threshold temperature $T_{th}$ as a function of
the magnetic field for several values of XY-anisotropy $\gamma$.
The curves mean the threshold temperature, limiting entangled (E)
and disentangled (D) region as a function of the magnetic field, assuming
fixed parameters values $J_{0}/J=-0.3$ and $J_{z}/J$=0.3, for several
values of XY-anisotropy $\gamma$. Fig. \ref{fig:Threshold}(a) for
$\gamma=0$, is observed the threshold temperature upper region corresponds
to entangled (E) and bottom region corresponds to disentangled (D)
region. (b) For $\gamma=0.1$, we observe a rise of threshold temperature
$T_{th}$ for the left side curve. From fig.\ref{fig:Threshold}(a-b), we observe those threshold temperature curves are highly sensitive for $\gamma\ne 0$, because the entanglement are extremely small close to threshold temperature. 
 (c) For $\gamma=0.4$, the reentrance
temperature decreases and also rise a small disentangled region below
entangled region for $h/J\lesssim0.6$. (d) For $\gamma=0.5$, the
disentangled region increases and still remains the reentrance temperature
$T_{th}$ for the right side curve. (e) For $\gamma=0.53$, the entangled
region shrinks even more and we can observe two quite interesting
reentrance temperature. (f) Increasing little bit the XY-anisotropy
$\gamma=0.535$, we observe how the reentrance temperature modify.
(g) Despite being similar to the previous plot for $\gamma=0.56$,
is illustrated how the reentrance temperature vanishes. (h) For $\gamma=0.585$,
the threshold temperature behaves similarly to the previous plot (g).
(i) For $\gamma=0.6$ the typical reentrance temperature vanishes,
despite there is a small isolated disentangled region.

\section{Conclusions}

In summary, we have presented a detailed study of the spin-1/2 Ising-XYZ
chain on diamond structure, which is an exactly solvable model through the decoration transformation and transfer matrix approach.
The phase diagram of ground state energy  was discussed, displaying
frustrated region and  the pairwise thermal entanglement between
two Heisenberg spin. It is noteworthy that  at
zero temperature, there are non-zero concurrence in a vast region
for the Hamiltonian parameter, only in the limiting case the entanglement
vanishes for some regions such as the case of  Ising-XXZ diamond chain\cite{spra}. Due to XY-anisotropy  the Hamiltonian, we
found quite interesting behavior for the thermal entanglement, such
as the presence of more than one threshold temperature, this phenomenon
is influenced by XY-anisotropy ($\gamma$), which means the entanglement vanishes at threshold temperature and for higher temperature arise the thermal entanglement again, and finally, for an even higher temperature, the entanglement vanishes definitely.
It is worth to emphasize, a similar result
was discussed in reference\cite{G. Rigolin,Zhou-Song}, but for only two qubits,
and here we propose how this property can be possible measured in
condensed matter materials.

\section*{Acknowledgment}

J. Torrico and M. Rojas thanks CAPES for fully financial support.
O. R. and de Souza thank CNPq and Fapemig for partial financial support.
This work was partly supported by the Brazilian FAPEMIG CEX - BPV
- 00046-13 and SCS MES RA in the frame of the research project No.
SCS 13-1C137 grants (N. A.).

\end{document}